\documentclass[12pt, draftclsnofoot, onecolumn]{IEEEtran}
\usepackage{graphicx,amsmath,amssymb}
\usepackage{subfigure}
\usepackage{citesort}
\usepackage{fancyhdr}
\usepackage{mdwmath}
\usepackage{mdwtab}
\usepackage{balance}
\usepackage{xcolor}
\usepackage{bm}
\usepackage{amsthm}
\usepackage{algorithm}
\usepackage{algorithmic}
\usepackage{multirow}
\usepackage{flafter}
\usepackage{setspace}
\usepackage{cite}
\usepackage{epstopdf}

\newtheorem{remark}{Remark}

\newtheorem{theorem}{Theorem}

\newtheorem{lemma}{Lemma}

\newtheorem{corollary}{Corollary}

\newtheorem{assumption}{Assumption}

\begin{document}
\title{Reconfigurable Intelligent Surfaces Aided Multi-Cell NOMA Networks: A Stochastic Geometry Model}

\author{Chao~Zhang,~\IEEEmembership{Student Member,~IEEE,}
        Wenqiang~Yi,~\IEEEmembership{Student Member,~IEEE,}
        and Yuanwei~Liu,~\IEEEmembership{Senior Member,~IEEE}
\thanks{C. Zhang, W. Yi, and Y. Liu are with Queen Mary University of London, London, UK (email:\{chao.zhang, w.yi, yuanwei.liu\}@qmul.ac.uk).}
}

\maketitle
\begin{abstract}
  By activating blocked users and altering successive interference cancellation (SIC) sequences, reconfigurable intelligent surfaces (RISs) become promising for enhancing non-orthogonal multiple access (NOMA) systems. The downlink RIS-aided NOMA networks are investigated via stochastic geometry. We first introduce the unique path loss model for RIS reflecting channels. Then, we evaluate the angle distributions based on a Poisson cluster process (PCP) model, which theoretically demonstrates that the angles of incidence and reflection are uniformly distributed. Additionally, we derive closed-form analytical and asymptotic expressions for coverage probabilities of the paired NOMA users. Lastly, we derive the analytical expressions of the ergodic rate for both of the paired NOMA users and calculate the asymptotic expressions for the typical user. The analytical results indicate that 1) the achievable rates reach an upper limit when the length of RIS increases; 2) exploiting RISs can enhance the path loss intercept to improve the performance without influencing the bandwidth. Our results show that 1) RIS-aided NOMA networks have superior performance than the conventional NOMA networks, and 2) the SIC order in NOMA systems can be altered since RISs are able to change the channel quality of NOMA users.
\end{abstract}

\begin{IEEEkeywords}
{M}ulti-cell NOMA, Poisson cluster process, reconfigurable intelligent surface
\end{IEEEkeywords}
\vspace{3cm}
\section{Introduction}

Due to introducing new freedom, non-orthogonal multiple access (NOMA) evolves into a promising technique. By sharing spectrum with power multiplexing schemes and successive interference cancellation (SIC), the spectral efficiency and user connectivity enable to be significantly improved to satisfy different target requirements~\cite{YuanweiNOMA}. In spite of benefits, NOMA techniques still have several significant implementation challenges such as lower received power and severer interference than orthogonal multiple access (OMA) users. To cope with the challenges, reconfigurable intelligent surfaces (RIS), also known as intelligent reflecting surfaces (IRS), hold promise in several aspects \cite{9086766,8910627}. On the one hand, environmental obstacles may ruin the required channel condition, especially for the far NOMA users blocked by high buildings, which results in inevitable outage situations. With the aid of RIS-introduced line-of-sight (LOS) propagation, we are able to exploit reflecting links through RISs to improve the channel condition of blocked NOMA users~\cite{8910627}. On the other hand, RISs are enabled to achieve flexible decoding orders according to the quality of service (QoS) conditions.

A RIS is regarded as a two-dimensional-equivalent reconfigurable meta-material, which is consist of elementary elements called scattering particles or meta-atoms \cite{RISURSorRIS,Passlossmodel}. Based on intelligent meta-surface technologies, RISs have properties such as absorbing incident waves or modifying the reflected wavefronts \cite{wirelesscom,pan}. In contrast to mirrors, RISs are able to adjust the angle of reflection and electric field strength. A major and basic open research challenge is to investigate the path loss model of RIS reflecting channels. Recent research contributions have studied the path loss model based on two typical methods, which are 1) correlated to the sum of incidence and reflection distances, and 2) correlated to the product of these distances. According to a fundamental work \cite{RISURSorRIS}, both of the typical methods are correct but utilizing in different application scenarios such that: 1) ``sum of distances'' model is suitable for short-distance communications such as indoor scenarios, while 2) ``product of distances'' model is suitable for long-distance communications such as outdoor scenarios. To reduce the path loss and interference, RISs are placed near to the served NOMA user. This spatial grouping property is depicted by a tractable stochastic geometry model, namely the Poisson cluster process (PCP)~\cite{7982794,8856258,GammaDistributionTransfer,haenggi}, which provides a theoretical framework for investigating the average performance of RIS-aided NOMA networks.

\subsection{Related Works}

\subsubsection{Related Works for RIS}
Sparked by the aforementioned potential benefits of RISs, recent research contributions on RIS-aided networks have been evaluated in various aspects. Firstly, one promising topic is to propose the passive beamforming designs, i.e., passive beamforming and information transfer techniques with a SAA-based iterative algorithm and a turbo message passing algorithm ~\cite{Beamforming1}, passive beamforming with modulation and resource allocation~\cite{Beamforming2}, and the achievable rates where limited discrete phase shifts are realized by the RIS \cite{Beamforming3}. Additionally, with various optimization designs, the performance of the RIS-aided networks is significantly enhanced: 1) the reflecting and precoding designs to minimize the symbol error rate were evaluated~\cite{OP1}; 2) the weighted sum rate by jointly designing the beamforming and the phase vector of the RIS was maximized \cite{OP2}; 3) the asymptotic results of the maximum and minimum signal-to-interference-plus-noise-ratio (SINR) was investigated \cite{OP3}. Moreover, contributions on deep learning methods have been investigated via \cite{Deeplearning1,Deeplearning2} and other related works with the aid of RIS have been investigated in several scenarios, such as mmWave environments \cite{scenario1,scenario2}, Internet of Things networks \cite{scenario3}, and RIS-empowered multi-input-multi-output (MIMO) systems \cite{scenario4}.

\subsubsection{Related Works for RIS-aided NOMA}
Additionally, with the aid of the RIS, the amounts of applications on RIS-NOMA systems attract our attention \cite{sumrate,BER,power2,power1,power3}. More specifically, an uplink RIS-aided NOMA system has been investigated to maximize the sum-rate \cite{sumrate}. Analytical results of the bit error rate (BER) for the downlink RIS-aided NOMA systems have been analyzed \cite{BER}. Moreover, the optimization on power-efficient for RIS-aided NOMA systems has been considered in \cite{power2,power1,power3}, i.e., optimizing the beamforming vectors and the IRS phase shift matrix compared with zero-forcing beamforming~, comparing power efficient of NOMA and OMA cases with the aid RISs~\cite{power1}, and proposing a difference-of-convex algorithm and efficient user ordering scheme to minimize the power consumption~\cite{power3}. With respect to physical layer analysis on RIS-NOMA networks, several contributions have provided with the approaches to investigate the RIS-aided NOMA networks, such as several RIS-NOMA beamforming methods in \cite{zhiguo2} and deriving outage probability of a single-cell RIS-NOMA network in \cite{zhiguo1}. With the aid of former efforts, contributions on RIS-aided NOMA networks have been driven \cite{tianwei1,yanyu,tianwei2}. In detail, the outage probability and the ergodic rate for both downlink and uplink scenarios under the RIS-aided single-cell NOMA networks have been derived \cite{yanyu}. Additionally, the stochastic geometry models to evaluate the spacial effect under RIS-aided single-cell NOMA networks have been considered in \cite{tianwei1,tianwei2}.

\subsection{Motivation and Contributions}

With the aid of considerable amounts of the former valuable contributions to RIS-aided single-cell NOMA networks, we extend the RIS-aided model from single-cell scenarios to multi-cell scenarios to investigate the performance of practical cases. With the aid of stochastic geometry models, we introduce a general case of path loss models, followed by an application of the ``product of distances'' model to analyze multi-cell NOMA networks for outdoor scenarios via a PCP-based spatial model.

Since RISs are capable to emit signals from BSs to various directions, we motivate to exploit RIS to enhance the channel quality of blocked users. Additionally, since the channel quality of RIS-aided channels is better than the non-RIS-aided channels, we are able to alter the SIC order via RISs to avoiding the delay-sensitive near users to accomplish SIC procedures. Motivated by the aforementioned challenges, we investigate RIS-aided multi-cell NOMA networks. The main contributions are summarized as follows:
 \begin{itemize}
 \item We express a general model of RIS reflecting links. Based on the general case of RIS reflecting links, we derive the path loss model in long-distance regions. By modeling the multi-cell networks as a PCP distribution, we investigate the angle distributions, which verify that the angles formed by users, RISs and base stations (BSs) are uniformly distributed in $[0,\pi]$.
  \item We derive closed-form analytical expressions for coverage probabilities of the paired NOMA users to enhance the evaluation efficiency. To investigate the impact of RISs, we calculate the asymptotic expressions for the typical user versus the half-length of RISs $L$. The analytical results indicate that we are capable of improving the coverage probability by adding the length of RISs.
  \item We derive the closed-form analytical expressions for ergodic rates of the paired NOMA users. Additionally, we calculate the asymptotic ergodic rate expressions for the typical user versus the half-length of RISs. The analytical results show that the achievable rates reach an upper limit when continuously enhancing the length of RISs. Additionally, enlarging the length of RISs can improve the path loss intercept, thereby enhance the coverage performance.
  \item Numerical results illustrate the following conclusions. 1) For the connected user, RIS-aided NOMA channels have superior coverage performance than RIS-aided OMA channels while it is on the contrary for the typical user. Additionally, for both NOMA users, RIS-aided NOMA networks acquire significantly enhanced performance than conventional NOMA scenarios. 2) If near users are delay-sensitive users, we enable to exploit the RISs to avoid the SIC procedures at near users. This is because the RISs have the ability to enhance the channel quality effectively to alter the SIC orders.
\end{itemize}

\subsection{Organizations}

The remaining sections of this paper are organized as follows. In Section II, the system model of this RIS-aided multi-cell NOMA networks is introduced, including the path loss model of RIS reflecting links and signal models. In Section III, we derive the path loss models for the long-distance communication scenarios, including the analysis of the angle and distance distributions. In Section IV, we derive the closed-form analytical expressions of coverage probability for the paired NOMA users. Additionally, we derive the asymptotic expressions of coverage probability for the typical user versus the length of RIS to evaluate the impact of RIS. In section V, we investigate the ergodic rate performance with closed-form analytical and asymptotic expressions. Numerical results are indicated in Section VI, followed by the conclusions in Section VII.

\section{System Model}

This paper considers RIS-aided downlink NOMA networks, where BSs and users are modeled according to two independent homogeneous Poisson point processes (HPPPs), namely $\Phi_b\subset \mathbb{R}^2$ with density $\lambda_b$ and $\Phi_u\subset \mathbb{R}^2$ with density $\lambda_u$, respectively. Two-user NOMA groups are served by orthogonal frequencies to cancel intra-cell interference. In each group, we assume that one of the paired users has already been connected to a BS in the previous user association process~\cite{7982794}. The other one, namely the typical user, joins this occupied resource block by applying power-domain NOMA techniques. To simplify the analysis, the connected user is not included in the user set $\Phi_u$ and the distance between this user to its BS is invariable as $r_c$. The typical user is randomly selected from $\Phi_u$ and its location is fixed at the origin of the considered plane.

\subsection{LOS Ball Model}
The blockage model is significant for RIS-aided networks as one promising application of RISs is to enhance the performance of blocked users by providing LoS transmission~\cite{8910627}. We consider a LoS ball model in this work~\cite{8401954,GammaDistributionTransfer}. For the typical user, its LoS ball has a radius $R_L$. The transmitters within this ball provide LoS transmission, while those outside this ball have NLoS links. The LoS ball region of the typical user is given by $\mathbb{O}(0,R_L) \subset \mathbb{R}^2$, where $\mathbb{O}(a,b)$ represents an annulus with the inner radius $a$ and outer radius $b$. To ensure the RU link is LoS, the RIS is uniform distributed in the LOS ball area of the typical user. Due to considering a blocked typical user, the region of considered BSs is in the range $\mathbb{O}(R_L,\infty)$\footnote{For BSs located in the RIS ball area $\mathbb{O}(0,R_L)$, RISs may weaken their direct LoS transmission due to phase difference~\cite{8910627}. Coherent transmission is desired for this case, which is beyond the scope of this paper.}.

\subsection{RIS-aided Link Model}
We assume one RIS is employed for helping the typical user. Based on stochastic geometry principles and the randomness of the typical user, users and their serving RISs can be regarded as the Matern cluster process (MCP) pattern of PCP models with a fixed number of nodes in each cluster. More specifically, the possible typical users are the parent point process deployed by a HPPP, where we choose one of them as the considered typical user. The RISs are uniformly deployed in the clusters (LOS balls) as the daughter point process. The channel conditions of the connected user have been known with a fixed distance. Based on the MCP model, there are three significant communication links in the considered NOMA group: 1) BU link, the link between the typical user and its BS; 2) BR link, the link between the BS and the employed RIS; and 3) RU link, the link between the RIS and the typical user. This work focuses on analyzing a blocked typical user and the RIS is applied to establish LoS route between the typical user and BSs~\cite{8910627}. Therefore, the BU link is assumed to be NLoS and the BR and RU links are LoS. Moreover, all NLoS communications are ignored in this paper due to their negligible received power.

The association criterion for the typical user is to associate with the BS offering the highest received power, which means that the distance between the RIS and the associated BS is the nearest. We assume the locations of the RIS and the associated BS are $\mathbf{x}_R$ and $\mathbf{x}_B$. Therefore, the distance between the associated BS and the RIS is correspondingly expressed as
 \begin{align}
 \mathbf{x}_{BR} = \arg \min_{\mathbf{x}_B \in \Phi_{B}} \|\mathbf{x}_B-\mathbf{x}_R \|
 \end{align}
 where $\mathbf{x}_R \in \mathbb{O}(0,R_L)$, $\Phi_{B} \subset\Phi_b$, $\Phi_{B} \subset \mathbb{O}(R_L,\infty)$ and the location of arbitrary interfering BS is denoted by $\mathbf{x}_I \in \Phi_{r} \setminus  \mathbf{x}_B$.

\subsection{Path Loss Model}

\begin{figure}[!htb]
\centering
\includegraphics[width= 5in]{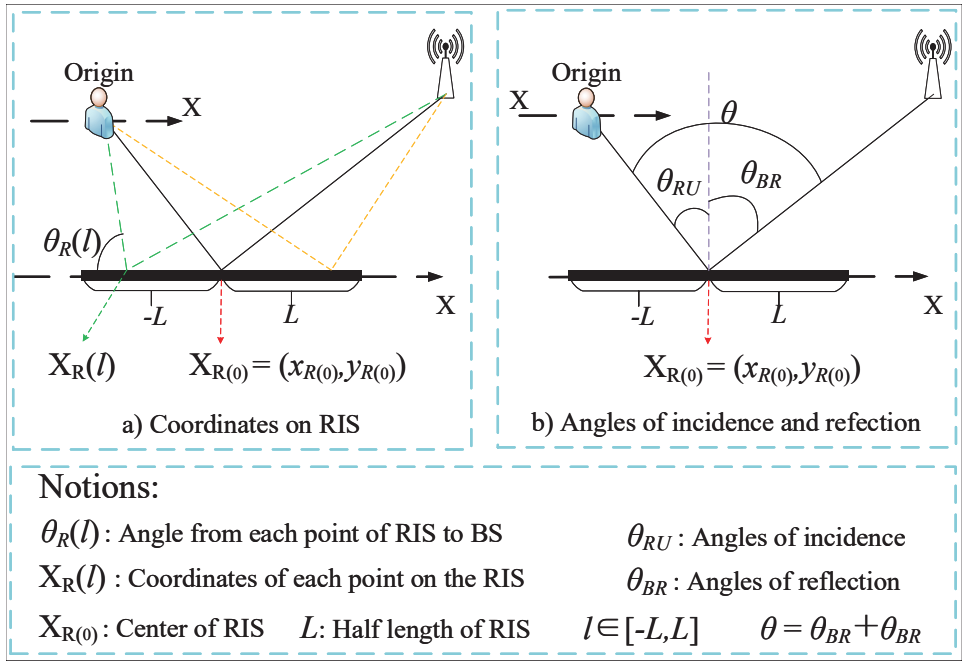}
\caption{Coordinates and angles in LOS balls: a) Coordinates on RIS; b) Angles of incidence and refection}
\label{angle}
\end{figure}

 This work considers a $2L$ linear RIS, whose central point is located at $\mathbf{X}_{R(0)}=\left(x_{R(0)},y_{R(0)}\right)\subset \mathbb{O}(0,R_L)$. We assume the associated BS and the typical user are distributed in the same side of the RIS to establish reflecting transmission~\cite{di2019reflection}. As shown in Fig. \ref{angle}, we set the x-axis is along the direction of RISs and the typical user is at the origin. Thus, the coordinate of each point on the RIS with a distance $l \in[-L,L]$ is expressed as $\mathbf{x}_R(l)=\left(x_R(0)+l\cos(\theta_R(l)),y_R(0)+l\sin(\theta_R(l))\right)\subset \mathbb{R}^2$, where $\theta_R(l)$ is the angle to x-axis for each point on the RIS. We set $r_{Q(0)}$ and $\theta_{Q(0)}$ as the distance and angle through the center of RISs $\mathbf{X}_{R(0)}$, where $Q \in \{BR, RU\}$ to represent the links from BSs to RISs and other links from RISs to the typical user. Additionally, we assume surfaces are deployed based on long-distance communication models, thereby the surfaces receive directional lights with approximations as ${r_Q}\left( x \right) \approx {r_{Q(0)}}+ qx\sin \left( {{\theta _{Q(0)}}} \right)$ where $q=1$ if $Q=BR$ and $q=-1$ if $Q=RU$~\cite{RISURSorRIS}.

 Under a high-frequency case with cylindrical electromagnetic (EM) waves~\cite{ntontin2019reconfigurable}, if one transmitting BS is located at $\mathbf{x}_b\in\{\mathbf{x}_B, \mathbf{x}_I\}$, the path loss model for the typical user is given by~\cite{RISURSorRIS}
 \begin{align}\label{P_t}
 P_t(\mathbf{x}_b,\mathbf{x}_R) = \left|\int_{ - L}^{ + L} {\Psi\left( l \right)} \exp \left( { - jk\Omega\left( l \right)} \right)dl\right|^2,
 \end{align}
 where
 \begin{align}
 &\Psi\left( x \right) = \frac{ {\cos \left( {{\theta _{\mathrm{BR}}(l)}} \right) + \cos \left( {{\theta _{\mathrm{RU}}(l)}} \right)} }{{8\pi {\sqrt{ {{r_{\mathrm{BR}}(l)}{r_{\mathrm{RU}}(l)}} }}}},\\
 &\Omega\left( x \right) = {r_{\mathrm{BR}}}\left( l \right) + {r_{\mathrm{RU}}}\left( l \right) - \Theta \left( l \right).
\end{align}
  The communication distance for the BR and the RU links are $r_{\mathrm{BR}} = \|\mathbf{x}_b-\mathbf{x}_R(l)\|$ and $r_{\mathrm{RU}} = \|\mathbf{x}_R(l)\|$, respectively. Considering the reflecting point is at $\mathbf{x}_R(l)$, the angle of incidence $\theta _{\mathrm{BR}}(l)\in\left(0,\frac{\pi}{2}\right]$ represents the angle between the corresponding BR link and the perpendicular bisector of the RIS, whilst the angle of reflection $\theta _{\mathrm{RU}}(l)\in\left(0,\frac{\pi}{2}\right]$ is the angle between the corresponding RU link and the perpendicular bisector of the RIS. The $\Theta \left( l \right)$ is the phase-shifting parameter of RISs which is decided by the desired transmitter and receiver.

 For the connected user at $\mathbf{x}_c \subset \mathbb{R}^2$, we assume the transmission between this user and its BS follows conventional wireless communications. Therefore, the path loss expression of the connected user is as follows
 \begin{align}
  P_c(\mathbf{x}_b,\mathbf{x}_c) = C\|\mathbf{x}_b-\mathbf{x}_c\|^{-\alpha_c},
 \end{align}
 where the $C$ is the intercept and $\alpha_c$ is the path loss exponent for the direct link. Note that the distance between the connected user and the associated BS is fixed. Therefore, $r_c = \|\mathbf{x}_B-\mathbf{x}_c\|$ is a constant.

 \begin{figure*}[t]
\centering
\includegraphics[width= 6.5in]{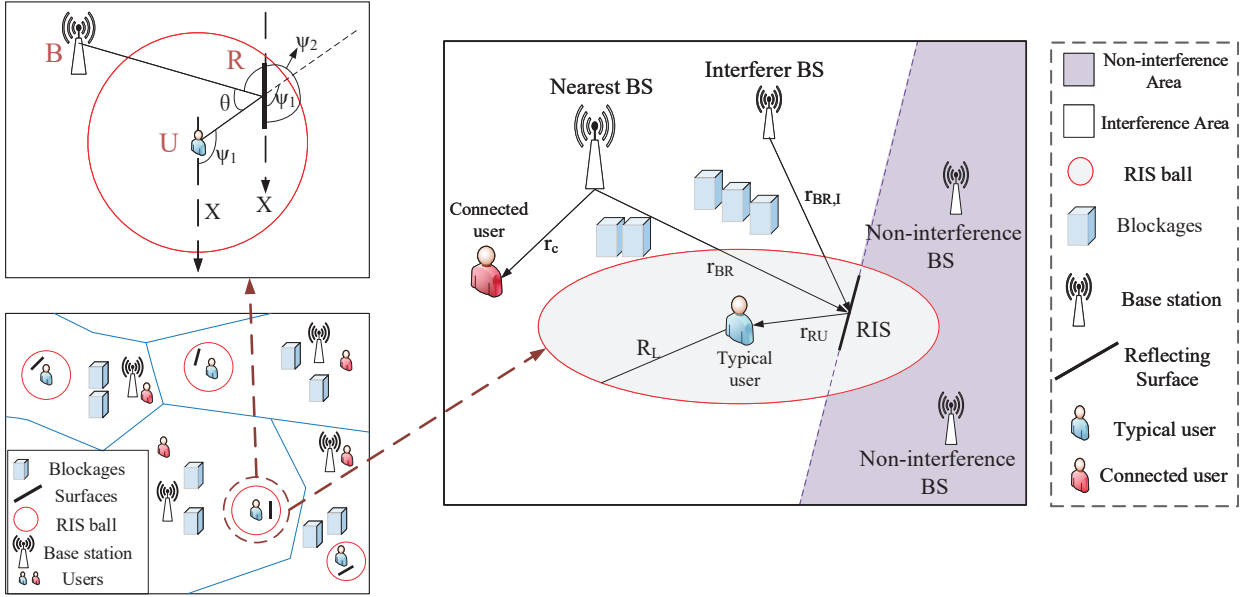}
\caption{Illustration of the signal model: (a) Bottom-left: An illustration of RIS-aided multi-cell scenarios; (b) Top-left: A typical NOMA network with angle demonstrations; (c) Right: A typical NOMA network with two types of BSs, such as interferer BSs facing RISs and Non-interference BSs against RISs.}
\label{fig_1}
\end{figure*}
 \subsection{Signal Model}
 In order to guarantee the quality of service (QoS) of the connected user, we assume the associated BS allocates more transmit power to the connected user than the typical user and SIC is processed at the typical user. Under this case, the typical user is regarded as transparent for the connected user. Therefore, the SINR for the SIC process at the typical user is given by
  \begin{align}
{\gamma _\mathrm{SIC}} = \frac{{{a_c}{P_b}{{\left| {{h_{t,\mathbf{x}_B}}} \right|}^2}P_c(\mathbf{x}_B,\mathbf{x}_c)}}{{ {{a_t}{P_b}{{\left| {{h_{t,\mathbf{x}_B}}} \right|}^2}P_t(\mathbf{x}_B,\mathbf{x}_R)} +I_{t,\rho_t} + {\sigma ^2}}},
 \end{align}
 where
 \begin{align}
 I_{t,\rho_t}=\rho_t\sum\limits_{\mathbf{x}_I \in \Phi_{r} \setminus  \mathbf{x}_B} {{P_b}{{\left| {{h_{t,\mathbf{x}_I}}} \right|}^2}P_t(\mathbf{x}_I,\mathbf{x}_R)}
 \end{align}
 and $P_b$ is the transmit power of BSs in each NOMA group and $\sigma ^2$ is the variance of additive white Gaussian noise (AWGN). The $a_t$ and $a_c$ are the power allocation parameters for the typical user and the connected user, respectively. Moreover, $a_c > a_t$ and $a_c + a_t = 1$. When the transmitter is at $\mathbf{x}$, for the receiver $\kappa$, the ${h_{\kappa,\mathbf{x}}}$ represents its Nakagami fading term with an integer parameter $m_\kappa$~\cite{8876629}. Additionally, $\kappa = c$ means the receiver is the connected user and $\kappa = t$ means that is the typical user. Regarding the interference $I_{t,\rho_t}$, since the signal from the back of RISs cannot pass through RISs, we assume $\rho_t \in [0,1]$ of the entire interference is able to reach the receiver $\kappa$.

 After the SIC process, the typical user decodes its data. The decoding SINR is expressed as
\begin{align}
{\gamma _t} = \frac{{{a_t}{P_b}{{\left| {{h_{t,\mathbf{x}_B}}} \right|}^2}P_t(\mathbf{x}_B,\mathbf{x}_R)} }{{I_{t,\rho_t} + {\sigma ^2}}}.
\end{align}

For the connected user, it directly decodes its messages by regarding the partner's signal as interference. Therefore, the decoding SINR for the connected user is as follows
\begin{align}
{\gamma _c} = \frac{{{a_c}{P_b}{{\left| {{h_{c,\mathbf{x}_B}}} \right|}^2}P_c(\mathbf{x}_B,\mathbf{x}_c)}}{{ {{a_t}{P_b}{{\left| {{h_{c,\mathbf{x}_B}}} \right|}^2}P_t(\mathbf{x}_B,\mathbf{x}_R)} +I_{c} + {\sigma ^2}}},
 \end{align}
 where
  \begin{align}
 I_{c}=\sum\limits_{\mathbf{x}_I \in \Phi_{r} \setminus  \mathbf{x}_B} {{P_b}{{\left| {{h_{t,\mathbf{x}_I}}} \right|}^2}P_c(\mathbf{x}_I,\mathbf{x}_c)}.
 \end{align}

It is worth noting that the connected user is interfered by all BSs excepting the associated BS.

\section{Path loss Analysis}

In this section, we derive the expressions of the path loss model and angle distributions, which are fundamental for analyzing the performance of RIS-aided NOMA systems.

\subsection{Path Loss Model}

Since surfaces are always installed on the walls of buildings, adjusting angles of surface physically may have constraints based on the shapes or directions of the walls. In this scenario, we operate the surfaces as anomalous reflectors, which is configured for reflecting waves towards a distinct direction of users. Hence, the reflection angle of waves is able to be unequal to the incidence angle.

\begin{lemma}\label{lemma2}
\emph{We regard surfaces as RISs rather than mirrors, thereby the angle of incidence enables to be different from the angle of reflection. Under this case, we are able to obtain the phase-shifting parameter as $\Theta \left( l \right) = \left( {\sin \left( {{\theta _{BR\left( 0 \right)}}} \right) - \sin \left( {{\theta _{RU\left( 0 \right)}}} \right)} \right)l + {{{\phi _0}}}/{k}$, with the aid of which, the direction is altered to the typical user via the phase shifters. Considered on the long-distance regions with directional lights, the path loss model on RIS operations is expressed as}
\begin{align}\label{PtRIS}
P_t^{RIS} \approx {C ^2_{RIS}}{\left( {{r_{{B}R\left( 0 \right)}}{r_{{R}U\left( 0 \right)}}} \right)^{ - {{{\alpha _t}}}}},
\end{align}
\emph{where ${C _{RIS}} = \frac{L}{{4\pi }}\left( {\cos \left( {{\theta _{{B}R\left( 0 \right)}}} \right) + \cos \left( {{\theta _{RU\left( 0 \right)}}} \right)} \right)$ and $\alpha_t$ is the path loss exponent of the typical user.}
\begin{IEEEproof}
Substituting $\Theta \left( x \right)$ into $\Psi\left( x \right)$, it is simplified as $\Psi\left( x \right) ={r_{{B}R\left( 0 \right)}} + {r_{RU\left( 0 \right)}} - \frac{{{\phi _0}}}{k}$. Since the BSs are assumed to emit cylindrical waves, we have assumptions such as ${r_Q}\left( x \right) \approx {r_{Q(0)}}+ qx\sin \left( {{\theta _{Q(0)}}} \right)$. Thus, the pass loss model is approximated as
\begin{align}
P_t^{RIS} &\approx {\left| {\frac{L}{{{\rm{4}}\pi }}{{\left( {{r_{{B}R}}\left( l \right){r_{RU}}\left( l \right)} \right)}^{ - \frac{{{\alpha _t}}}{2}}}} \right|^2} {\left| {\left( {\cos \left( {{\theta _{{B}R\left( 0 \right)}}} \right) + \cos \left( {{\theta _{RU\left( 0 \right)}}} \right)} \right)} \right|^2} \notag \\
 &\hspace*{0.3cm} \times {\left| {\exp \left( { - jk\left( {{r_{{B}R\left( 0 \right)}} + {r_{RU\left( 0 \right)}} - \frac{{{\phi _0}}}{k}} \right)} \right)} \right|^2},
\end{align}
and we obtain \eqref{PtRIS} via algebraic manipulations.
\end{IEEEproof}
\end{lemma}

\subsection{Distance Distributions}

Note that the users and BSs are settled via two independent HPPPs and reflecting surfaces are uniformly deployed within the ball $\mathbb{O}(0,R_L)$ of the typical user. Based on the MCP pattern of PCP models, locations are defined that users are parent nodes obeying HPPP and surfaces are daughter nodes within the clusters of RIS balls. Based on the aforementioned settings, we derive the probability density functions (PDFs) of distances of the corresponding cluster and other clusters for a typical user.

\subsubsection{The Corresponding Cluster of the Typical User}
We focus on a typical user located at the center of the RIS ball area served by uniformly distributed intelligent surfaces \cite{yuanwei1}. Thus, we are capable of deriving the PDF of the distance from a surface to its targeted typical user, denoted as $r_{RU}$, as
\begin{align}\label{PDFRU}
{f_{{r_{RU}}}}\left( x \right) = \frac{{2x}}{{R_L^2}}U\left( {{R_L} - x} \right),
\end{align}
where $U\left( \cdot \right)$ is the unit step function.

\subsubsection{Other Clusters of the Typical User}
Since the LOS links from BSs to the typical user are blocked, we only investigate the reflecting links from the BSs to the surfaces. Thus, based on the null probability of a 2-D PPP within in the RIS ball area and order statistics \cite{nNearestUser, nNearestUserJ}, the PDF of the distance between a RIS and its $n^{th}$ nearest BS is derived as
\begin{align}\label{r_BU}
{f_{{r_{BR}}}}\left( {x,n} \right) = \frac{{2{{\left( {\pi {\lambda _b}} \right)}^n}}}{{\left( {n - 1} \right)!}}{x^{2n - 1}}\exp \left( { - \pi {\lambda _b}{x^2}} \right).
\end{align}

\subsection{Angle Distributions}

Shown as Fig. \ref{fig_1}, we denote a BS as node $B$, a RIS as node $R$, and a typical user as node $U$ to clarify the angles. With the aid of a chosen positive X-axis that is parallel to the RIS, the angles are illustrated as $\psi_1 = \angle{RUX}  $, $\psi_2 = \angle{BRX}$ and $\theta = \left|\pi-\left|\psi_2-\psi_1\right|\right|$. Notice that the angle of $\psi_2$ is uniformly distributed within $(0,2\pi)$ based on the properties of HPPP. We additionally observe that the angle of $\psi_1$ obeys uniform distribution in $(0,2\pi)$ since the RIS is uniformly distributed in the RIS ball. Based on $\psi_1$ and $\psi_2$ with the same distributions, the cumulative distribution function (CDF) of $ \left|\psi_2-\psi_1\right|$ is derived as
\begin{align}
{F_{\left| {{\psi _{\rm{2}}} - {\psi _1}} \right|}}\left( z \right)  &= \frac{{4\pi z - {z^2}}}{{4{\pi ^2}}},
\end{align}
therefore, the PDF of the angle of $ \left|\psi_2-\psi_1\right|$ is derived as ${f_{\left| {{\psi _{\rm{2}}} - {\psi _1}} \right|}}\left( z \right) = \frac{{2\pi  - z}}{{2{\pi ^2}}}$.

With the respect to $\theta = \left|\pi-\left|\psi_2-\psi_1\right|\right|$, the CDF of the angle $\theta$ is derived as
\begin{align}
&{F_\theta }\left( x \right) = {F_{\left| {{\psi _{\rm{2}}} - {\psi _1}} \right|}}\left( {x + \pi } \right) - {F_{\left| {{\psi _{\rm{2}}} - {\psi _1}} \right|}}\left( {x - \pi } \right) = \frac{x}{\pi },
\end{align}
which is proved that the angle $\theta$ obeys uniform distribution within $(0,\pi)$ with the PDF as $f_\theta (x) = 1/\pi$.

Recall that we denote the angles of incidence as $\theta _{BR\left( 0 \right)}$ and the angles of reflection as $\theta _{RU\left( 0 \right)}$, thereby we are able to observe $\theta = \theta _{BR\left( 0 \right)}+\theta _{RU\left( 0 \right)}$ from Fig. \ref{fig_1}. In the following, the angle analysis with respect to RISs is investigated.

When the surfaces are designed as RISs, the angles of incidence and reflection are unequal. We define the angles of incidence $\theta _{BR\left( 0 \right)}=\rho_a \theta$, where $\rho_a \in(0,1)$, thereby the angles of reflection is $\theta _{RU\left( 0 \right)} = (1-\rho_a)\theta$. Hence, the PDFs of the angles of incidence and reflection are derived as
\begin{align}
&\label{angle1}{f_{{\theta _{BR\left( 0 \right)}}}}(x) =  \frac{1}{{{\rho _a}\pi }},x \in \left( {0,\frac{\pi }{2}} \right),\\
&\label{angle2}{f_{{\theta _{RU\left( 0 \right)}}}}(x) = \frac{1}{{\left( {1 - {\rho _a}} \right)\pi }},x \in \left( {0,\frac{\pi }{2}} \right).
\end{align}

\begin{remark}\label{Angledistribution}
As shown in Fig. \ref{fig_1}, we note that the angles $\psi_1$ and $\psi_2$ are uniform distributed in the range of $[0,2\pi]$. Based on the derivations in this subsection, we conclude that the BS-RIS-User angle is uniformly distributed in $[0,\pi]$. In a word, we enable to extend the conclusion that any angle formed via three nodes of a PPP are uniformly distributed.
\end{remark}

\begin{lemma}\label{EverageAngle}
\emph{Considered on an average case with a 2-D HPPP, the angles are uniformly distributed. Since communication distances are independent of incidence and reflection angles, after considering the random spatial effect, the path loss of RIS links in \textbf{Lemma \ref{lemma2}} is rewritten as follows}
\begin{align}
P_t^{RIS} = {C _{RIS,E}}{\left( {{r_{{B}R\left( 0 \right)}}{r_{{R_i}U\left( 0 \right)}}} \right)^{ - {{{\alpha _t}}}}},
\end{align}
\emph{where ${C _{RIS,E}}  =\frac{{{L^2}}}{{16{\pi ^3}}}\left( {\pi  + \frac{{\sin \left( {2{\rho _a}\pi } \right)}}{{4{\rho _a} - 12\rho _a^2 + \rho _a^3}}} \right)$.}
\begin{IEEEproof}
The average path loss with respect to angles is expressed as $P_t^{RIS} = \mathbb{E}\left[ {C_{RIS,I}^2} \right]$ $\times{\left( {{r_{{B}R\left( 0 \right)}}{r_{{R_i}U\left( 0 \right)}}} \right)^{ - {{{\alpha _t}}}}}$, where $\mathbb{E}[\cdot]$ is calculating the expectation. By plugging \eqref{angle1} and \eqref{angle2}, this lemma is proved.
\end{IEEEproof}
\end{lemma}

Under a vacuum scenario with the best channel environment as mentioned in \cite{RISURSorRIS}, the path loss exponent of reflecting links $\alpha_t=1$. For practical scenarios, this path loss exponent $\alpha_t$ should be larger than one. Therefore, we consider $\alpha_t > 1$ in the rest of this work. The path loss exponent for conventional models are defined as $\alpha_{RF} >2$.

\begin{remark}\label{pathloss}
Note that the conventional path loss model for radio frequency (RF) models is expressed as ${P_{RF}} = {C_{RF}}d_{RF}^{ - {\alpha _{RF}}}$, where ${C_{RF}} = {\left( {\frac{c}{{4{f_c}\pi }}} \right)^2}=\frac{1}{{16{\pi ^2}}}{\left( {\frac{c}{{{f_c}}}} \right)^2}$ is the intercept, $c=3 \times 10^8$ m/s is the speed of light, $f_c$ is the bandwidth of users. We approximately express the intercept RIS models as ${C_{RIS,E}} = \frac{{{L^2}}}{{16{\pi ^2}}}$. Thus, we are capable to compare the intercepts by comparing $L$ and $c/f_c$. With the aim to enhance the performance by improving the intercepts, we need to enlarge the length $L$ for RIS scenarios and to reduce the bandwidth $f_c$ for RF scenarios. However, reducing bandwidth may cause the decline of performance, which is a contradiction from our aims. Thus, exploiting RISs is capable of optimizing the path loss model to improve the performance without influencing the bandwidth.
\end{remark}

\section{Coverage Performance Evaluation}

When we pre-decide a fixed threshold rate, the communication performance is guaranteed when the transmit rates are higher than the threshold. By defining a fixed threshold to evaluate whether the QoS of a network is satisfied, we investigate the SINR coverage performance on our RIS-aided NOMA networks based on the derived path loss expressions.

Since we exploit the typical user with SIC procedures and the messages connected user are directly decoded, the coverage probability expressions for the connected user and a typical use are expressed respectively as
\begin{align}
&\label{coverage_connected}{\mathbb{P}}_{t}={\rm{Pr}}\left\{ {{\gamma _{{\rm{SIC}}}} > \gamma _{SIC}^{th},{\gamma _t} > \gamma _t^{th}} \right\},\\
&\label{coverage_typical}\mathbb{P}_{c}={\rm{Pr}}\left\{ {{\gamma _{\rm{c}}} > \gamma _c^{th}} \right\},
\end{align}
where $\Pr\{\cdot\}$ is the probability operation, ${\gamma _{SIC}^{th}}$ is the threshold of SIC procedures, $\gamma _t^{th} = {2^{{R_t}/{B_w}}} - 1$ is the coverage threshold of the typical user with threshold rate $R_t$ and bandwidth $B_w$, $\gamma _c^{th} = {2^{{R_c}/{B_w}}} - 1$ is the threshold of the connected user with threshold rate $R_c$.

\subsection{Interference Analysis}
Before evaluating the coverage performance of this network, we would first derive the Laplace transform of interference, ${I_{t,{\rho _t}}}$ and $I_{c}$, under two scenarios.

\subsubsection{Interference Analysis of the Connected User}
Since the connected user is not served by reflecting surfaces, the Laplace transform of the interference for the connected user is expressed via conventional wireless communication analysis~\cite{yuanwei1} as
\begin{align}\label{I_C_def}
&{\cal L}_{(s)}^{c} =\mathbb{E} \left[ \exp\left(-{\sum\limits_{{{\bf{x}}_I} \in {\Phi _r} \setminus {{\bf{x}}_B}} {{P_b}C{{\left| {{h_{t,{{\bf{x}}_I}}}} \right|}^2}{r_{c,I}}^{ - {\alpha _c}}} }\right) \right].
\end{align}

\begin{lemma}\label{I_c}
\emph{The Laplace transform of interference for the connected user is derived as}
\begin{align}
{\cal L}_{(s)}^c = \exp \left( { - {\varsigma _1}\left( {{}_2{F_1}\left( { - \frac{2}{{{\alpha _c}}},m;1 - \frac{2}{{{\alpha _c}}}; - \varsigma_2 s} \right) - 1} \right)} \right),
\end{align}
\emph{where ${}_2{F_1}\left( \cdot,\cdot;\cdot; \cdot \right)$ is the hypergeometric function, ${\varsigma _1} = \pi {\lambda _b}r_c^2$ and ${\varsigma _2} = \frac{{{P_b}C}}{{mr_c^\alpha }}$.}
\begin{IEEEproof}
See Appendix~A.
\end{IEEEproof}
\end{lemma}

\subsubsection{Interference Analysis of the Typical on RIS Scenario}
With the aid of \textbf{Lemma \ref{EverageAngle}}, the Laplace transform of the interference under RIS scenarios is expressed as
\begin{align}
&{\cal L}_{(s)}^{t,RIS} = \mathbb{E}\left[ \exp\left(-{{\rho _t}\sum\limits_{{{\bf{x}}_I} \in {\Phi _r} \setminus {{\bf{x}}_B}} {\frac{{{P_b}C_{RIS,I }^2{{\left| {{h_{t,{{\bf{x}}_I}}}} \right|}^2}}}{{{{\left( {{r_{BR\left( I  \right)}}{r_{RU\left( I  \right)}}} \right)}^{{\alpha _t}}}}}} } \right)\right],
\end{align}

\begin{lemma}\label{I_t}
\emph{With the aid of RISs, the Laplace transform of interference for the typical user is derived as}
\begin{align}
 &{\cal L}_{(s)}^{t,RIS}\left( {{r_{BR\left( 0 \right)}},{r_{RU\left( 0 \right)}}} \right)= \exp \left( { - {\varsigma _3}\left( {{}_2{F_1}\left( { - \frac{2}{{{\alpha _t}}},m;1 - \frac{2}{{{\alpha _t}}}; - {{s{\varsigma _4}}}{{}}} \right) - 1} \right)} \right),
\end{align}
\emph{where ${\varsigma _3} = \pi {\lambda _b}r_{BR\left( 0 \right)}^2$ and ${\varsigma _4} = {P_b}C_{RIS,E}^2\frac{{1}}{{{m_t}{r_{RU\left( 0 \right)}}{r_{BR\left( 0 \right)}}^{{\alpha _t}}}}$.}
\begin{IEEEproof}
See Appendix~B.
\end{IEEEproof}
\end{lemma}

\subsection{Coverage Analysis with RISs}
In this subsection, with the aid of RISs, the closed-form expressions of the coverage probabilities for the typical user and the connected users are derived via \textbf{Theorem \ref{CPRIS}-\ref{CPC}} and \textbf{Corollary \ref{CPRIS1}-\ref{CPRIS2}}.

\subsubsection{Coverage Analysis of the Typical User with RISs}
Note that the interference from the typical user is strived to be canceled with the aid of SIC procedures. When the surfaces perform as RISs, based on \textbf{Lemma \ref{EverageAngle}} and \eqref{coverage_typical}, the coverage probability is rewritten as
\begin{align}
{\mathbb{P}}_{t} = {\rm{Pr}}\left\{ {{{\left| {{h_{t,{{\bf{x}}_B}}}} \right|}^2} > \frac{{\Upsilon \left( {{I_{t,{\rho _t}}} + {\sigma ^2}} \right)}}{{{P_b}P_t^{RIS}}}} \right\},
\end{align}
where $\Upsilon  = \max \left( {\frac{{\gamma _{SIC}^{th}}}{{{a_c} - \gamma _{SIC}^{th}{a_t}}},\frac{{\gamma _t^{th}}}{{{a_t}}}} \right)$.

\begin{theorem}\label{CPRIS}
\emph{Note that we assume reflecting channels as Nakagami-m fading channels. With the aid of RISs, the approximated expression of coverage probability for the typical user is derived as}
\begin{align}\label{CPRRISexact}
{\mathbb{P}}_{t} &\approx  2\pi {\lambda _b}\int_0^{{R_L}} {\int_0^\infty  {\sum\limits_{n = 1}^{{m_t}} {{{\left( { - 1} \right)}^{n + 1}} {\binom{m_t}{n}} } x} }\exp \left( { - {\beta _{\rm{0}}}\left( y \right){x^{{\alpha _t}}}} \right)\exp \left( { - {\beta _2}{x^2}} \right)dx{f_{{r_{RU}}}}\left( y \right)dy,
\end{align}
\emph{where ${\beta _{\rm{0}}}\left( x \right){\rm{ = }}{\beta _{\rm{1}}}x^\alpha_t$, ${\beta _{\rm{1}}}{\rm{ = }}\frac{{n{\eta _t}\Upsilon {\sigma ^2}}}{{{P_b}C_{RIS,E}^2}}$, ${\beta _2} = \pi {\lambda _b}{}_2{F_1}\left( { - \frac{2}{{{\alpha _t}}},m;1 - \frac{2}{{{\alpha _t}}}; - \frac{{n{\eta _t}\Upsilon }}{{{m_t}}}} \right)$ and $m_t$ is the coefficient in Nakagami-m fading channels with unit mean values.}
\begin{IEEEproof}
See Appendix~C.
\end{IEEEproof}
\end{theorem}

\begin{remark}\label{RT1}
When we analyze the performance with respect to the length of RISs $L$, we simplify the expression of coverage probability based on \textbf{Theorem \ref{CPRIS}} as
 \begin{align}
 {\mathbb{P}}_{t} &\approx \exp \left( { - \frac{{{A_1}\left( {x,y} \right)}}{{{L^2}}}} \right){A_2}\left( {x,y} \right).
 \end{align}
where ${{A_1}\left( {x,y} \right)}$ and ${A_2}\left( {x,y} \right)$ are other items irrelevant to the length of RISs $L$. Thus, we conclude that when we improve the length of RISs, the coverage performance is enhanced. This is because the large RISs are able to absorb more incident waves and reflect them to the users.
\end{remark}

\begin{corollary}\label{CPRIS1}
\emph{Conditioned on $\alpha_t = 2$, we are capable to derive the closed-form expression of the coverage probability for the typical user as}
\begin{align}
{\mathbb{P}}_{t} &\approx    \frac{{\pi {\lambda _b}}}{2}\sum\limits_{n = 1}^{{m_t}} {{{\left( { - 1} \right)}^{n + 1}} {\binom{{m_t}}{n}}} \left( {{\beta _{\rm{1}}}R_L^2{\rm{ + }}2{\beta _2}} \right).
\end{align}
\begin{IEEEproof}
When $\alpha_t=2$, the coverage probability of the typical user is rewritten as
\begin{align}\label{ApB1}
\mathbb{P}_c \approx  2\pi {\lambda _b}\int_0^{{R_L}} {\int_0^\infty  {\sum\limits_{n = 1}^{{m_t}} {{{\left( { - 1} \right)}^{n + 1}} {
\binom{{m_t}}{n}
} } x\exp \left( { - \left( {{\beta _{\rm{0}}}\left( y \right){\rm{ + }}{\beta _2}} \right){x^{\rm{2}}}} \right)} } dx{f_{{r_{RU\left( 0 \right)}}}}\left( y \right)dy,
\end{align}

With the aid of Gamma functions $\Gamma(1)=1$ and Eq. [2.3.18.2] in \cite{table}, \eqref{ApB1} is derived as
\begin{align}\label{ApB2}
\mathbb{P}_c \approx \pi {\lambda _b}\sum\limits_{n = 1}^{{m_t}} {{{\left( { - 1} \right)}^{n + 1}} {\binom{{m_t}}{n}} } \int_0^{{R_L}} {\left( {{\beta _{\rm{0}}}\left( y \right){\rm{ + }}{\beta _2}} \right)} {f_{{r_{RU\left( 0 \right)}}}}\left( y \right)dy,
\end{align}
and via several algebraic manipulations, the final expression is obtained.

\end{IEEEproof}
\end{corollary}

\begin{corollary}\label{CPRIS2}
\emph{Conditioned on $\alpha_t = 4$, the closed-form expression of the coverage probability for the typical user is derived via the error function as}
\begin{align}
{\mathbb{P}}_{t} &\approx  \sum\limits_{n = 1}^{{m_t}} {{{\left( { - 1} \right)}^{n + 1}}{\binom{m_t}{n}} } \sum\limits_{i = 1}^K {\frac{{{\omega _i}{\pi ^{\frac{3}{2}}}{\lambda _b}\sqrt {1 - \Xi _i^2} }}{{2R_L^{}\sqrt {{\beta _1}} {\Xi _i}}}}  \exp \left( {\frac{{\beta _2^2}}{{4{\beta _1}\Xi _i^4}}} \right)\rm{Erfc}\left( {\frac{{\beta _2^{}}}{{2\sqrt {{\beta _1}} \Xi _i^2}}} \right),
\end{align}
\emph{where ${\eta _t} = {m_t}{\left( {{m_t}!} \right)^{ - \frac{1}{{{m_t}}}}}$, ${\varpi _i}{\rm{ = cos}}\left( {\frac{{2i - 1}}{{2K}}\pi } \right)$, ${\Xi _i} = \frac{{{R_L}}}{2}\left( {{\varpi _i} + 1} \right)$, ${\omega _i} = \pi /K$ and $\rm{Erfc}(\cdot)$ is the error function.}
\begin{IEEEproof}
Based on Appendix~C when $\alpha_t=4$, this corollary is proved via substituting \eqref{PDFRU}, Eq. [2.3.15.4] in \cite{table} and Chebyshev-gauss quadrature into \eqref{CPRRISexact}.
\end{IEEEproof}
\end{corollary}

\subsubsection{Coverage Analysis of the Connected User}
Based on \eqref{coverage_connected}, we are capable of rewriting the coverage probability expression as
\begin{align}
{\mathbb{P}}_{c} =  &1 - {\rm{Pr}}\left\{ {{{\left| {{h_{c,{{\bf{x}}_B}}}} \right|}^2} < \frac{{\gamma _c^{th}\left( {{I_c} + {\sigma ^2}} \right)}}{{\left( {{a_c} - {a_t}\gamma _c^{th}} \right){P_b}C{r_c}^{ - {\alpha _c}}}}} \right\}.
\end{align}

\begin{theorem}\label{CPC}
\emph{The closed-form expression of coverage probability for the connected users is derived as}
\begin{align}\label{CPCexact}
{\mathbb{P}}_c &\approx   \sum\limits_{n = 1}^{{m_c}} {{{\left( { - 1} \right)}^{n + 1}} {
\binom{m_c}{n}
}} \exp \left( { - {\mu _3}r_c^2 - {\mu _4}{r_c}^{{\alpha _c}}} \right),
\end{align}
\emph{where ${\mu _1} = \pi {\lambda _b}\left( {{}_2{F_1}\left( { - \frac{2}{{{\alpha _c}}},m;1 - \frac{2}{{{\alpha _c}}}; - \frac{{n{\eta _c}\gamma _c^{th}}}{{m\left( {{a_c} - {a_t}\gamma _c^{th}} \right)}}} \right) - 1} \right)$, ${\mu _2} = \frac{{n{\eta _c}\gamma _c^{th}{\sigma ^2}}}{{\left( {{a_c} - {a_t}\gamma _c^{th}} \right){P_b}C}}$ and ${\eta _c} = {m_c}{\left( {{m_c}!} \right)^{ - \frac{1}{{{m_c}}}}}$ with Nakagami-m fading coefficient $m_c$.}
\begin{IEEEproof}
We exploit the Campbell¡¯s theorem and \textbf{Theorem \ref{CPRIS}} to derive \eqref{CPCexact}.
\end{IEEEproof}
\end{theorem}

\begin{remark}\label{RT2}
When we analyze the performance with respect to the density of BSs $\lambda_b$, we simplify the expression of coverage probability as
 \begin{align}
 {\mathbb{P}}_{t} &\approx \exp \left( { - {\lambda _b}{B_1}\left( {x,y} \right)} \right){B_2}\left( {x,y} \right)
 \end{align}
where ${{B_1}\left( {x,y} \right)}$ and ${B_2}\left( {x,y} \right)$ are other items irrelevant to the density of BSs $\lambda_b$. Hence, the expression reveals that when we improve the density of BSs $\lambda_b$, the coverage probability is reduced. This is because the distance of connected user has been fixed and the density of BSs $\lambda_b$ only influence the strength of interference from other BSs.
\end{remark}

\subsection{Asymptotic Coverage Probability for the Typical user}
In this subsection, we evaluate the coverage performance of the typical user when the half-length $L \to \infty$.

\begin{corollary}\label{co3}
\emph{Conditioned on $L \to \infty$, the asymptotic expression of coverage probability for the typical user is derived as }
\begin{align}
 {\mathbb{P}}_c &\approx \sum\limits_{n = 1}^{{m_t}} {{{\left( { - 1} \right)}^{n + 1}}
\binom{m_t}{n}
} \frac{{\pi {\lambda _b}}}{{{\beta _2}}} - \frac{{2\pi {\lambda _b}R_L^{{\alpha _t}}}}{{\left( {2 + {\alpha _t}} \right)}}\Gamma \left( {\frac{{{\alpha _t} + 2}}{2}} \right)\sum\limits_{n = 1}^{{m_t}} {{{\left( { - 1} \right)}^{n + 1}}
\binom{m_t}{n}
} \frac{{{\beta _1}}}{{\beta _2^{\frac{{{\alpha _t} + 2}}{2}}}}.
\end{align}
\begin{IEEEproof}
After substituting the asymptotic expression $\exp(-x)=1-x$ into the coverage probability expression, we are able to rewrite \eqref{CPRRISexact} as
\begin{align}
 {\mathbb{P}}_c &\approx  2\pi {\lambda _b}\int_0^{{R_L}} {\frac{{2y}}{{R_L^2}}dy\int_0^\infty  {\sum\limits_{n = 1}^{{m_t}} {{{\left( { - 1} \right)}^{n + 1}}\binom{m_t}{n}} x\exp \left( { - {\beta _2}{x^2}} \right)} } dx\notag\\
 &\hspace*{0.3cm}- 2\pi {\lambda _b}{\beta _1}\frac{{2R_L^{{\alpha _t}}}}{{\left( {2 + {\alpha _t}} \right)}}\int_0^\infty  {\sum\limits_{n = 1}^{{m_t}} {{{\left( { - 1} \right)}^{n + 1}}\binom{m_t}{n}} {x^{{\alpha _t} + 1}}\exp \left( { - {\beta _2}{x^2}} \right)},
\end{align}
and based on Eq. [2.3.18.2] in \cite{table}, we enable to obtain the final results.
\end{IEEEproof}
\end{corollary}

\begin{corollary}\label{R3}
When the length of RISs are sufficiently large, we are capable to derive an upper limit from \textbf{Corollary \ref{co3}} as
\begin{align}
 {\mathbb{P}}_c &\approx\sum\limits_{n = 1}^{{m_t}} {{{\left( { - 1} \right)}^{n + 1}}\binom{m_t}{n}} \frac{{\pi {\lambda _b}}}{{{\beta _2}}},
\end{align}
which can increase the calculation efficiency to a large extent.
\end{corollary}
\section{Ergodic Rate Evaluation}

Compared with the coverage probability with a fixed rate threshold, the achievable ergodic rate for the RIS-aided NOMA systems is opportunistically altered via the channel conditions of users. In this subsection, the closed-form expression of the ergodic rate for the typical user and the connected users are derived via \textbf{Theorem \ref{CPRIS}-\ref{CPC}} and \textbf{Corollary \ref{CPRIS1}-\ref{CPRIS2}}.

\subsection{Ergodic Rate for the Typical User}
We consider the typical user to exploit the SIC procedure. The failure SIC procedure leads to the ergodic rate of the typical user always being zero. Thus, with the aid of the expression of the coverage probability, we express the ergodic rate expression of the typical user as
\begin{align}\label{erdeft}
{\mathbb{E}}\left[ {R_t^{RIS}} \right]& = {\mathbb{E}}\left[ {{{\log }_2}\left( {1 + {\gamma _t}} \right),{\gamma _{{\rm{SIC}}}} > \gamma _{SIC}^{th}} \right]\notag\\
 &= \frac{1}{{\ln 2}}\int_0^\infty  {\frac{{{{\mathbb{P}}_t}\left( {\gamma _t^{th}} \right)}}{{1 + \gamma _t^{th}}}} d\gamma _t^{th},
\end{align}
and the approximated and closed-form expressions are derived in the following.

\begin{theorem}\label{ERRIS}
\emph{With the aid of the coverage probability expression in \textbf{Theorem \ref{CPRIS}}, the approximated expression of ergodic rates for the typical user is derived as}
\begin{align}
{{\mathbb{E}}}\left[ {R_t^{RIS}} \right] &= \frac{{2\pi {\lambda _b}}}{{\ln 2R_L^2}}\int_{{a_t}{\Upsilon _1}}^\infty  {\int_0^{{R_L}} {\int_0^\infty  {\sum\limits_{n = 1}^{{m_t}} {{{\left( { - 1} \right)}^{n + 1}} {
\binom{m_t}{n}
} } } } } \notag\\
 &\hspace*{0.3cm}\times \frac{{2yx}}{{1 + z}}\exp \left( { - {\beta _{\rm{1}}}\left( {\frac{z}{{{a_t}}}} \right){{\left( {yx} \right)}^{{\alpha _t}}}} \right)\exp \left( { - {\beta _2}\left( {\frac{z}{{{a_t}}}} \right){x^2}} \right)dxdydz\notag\\
& \hspace*{0.3cm}+ \frac{{2\pi {\lambda _b}}}{{\ln 2R_L^2}}\int_0^{{a_t}{\Upsilon _1}} {\int_0^{{R_L}} {\int_0^\infty  {\sum\limits_{n = 1}^{{m_t}} {{{\left( { - 1} \right)}^{n + 1}} {
\binom{m_c}{n}
} } } } }\notag \\
& \hspace*{0.3cm}\times \frac{{2yx}}{{1 + z}}\exp \left( { - {\beta _{\rm{1}}}\left( {{\Upsilon _1}} \right){{\left( {yx} \right)}^{{\alpha _t}}}} \right)\exp \left( { - {\beta _2}\left( {{\Upsilon _1}} \right){x^2}} \right)dxdydz,
\end{align}
\emph{where ${\beta _{\rm{1}}}\left( {\Upsilon \left( z \right)} \right) = \frac{{n{\eta _t}\Upsilon \left( z \right){\sigma ^2}}}{{{P_b}C_{RIS,E}^2}}$, $\Upsilon \left( z \right) = \max \left( {\frac{{\gamma _{SIC}^{th}}}{{{a_c} - \gamma _{SIC}^{th}{a_t}}},\frac{z}{{{a_t}}}} \right)$, ${\Upsilon _1} = {\frac{{\gamma _{SIC}^{th}}}{{{a_c} - {a_t}\gamma _{SIC}^{th}}}} $ and ${\beta _2}\left( z \right) = \pi {\lambda _b}{}_2{F_1}\left( { - \frac{2}{{{\alpha _t}}},m;1 - \frac{2}{{{\alpha _t}}}; - \frac{{n{\eta _t}\Upsilon \left( z \right)}}{{{m_t}}}} \right)$.}
\begin{IEEEproof}
When considering the threshold $\gamma _t^{th}$ as a variable, the condition to maintain the SIC orders as $\Upsilon \left( z \right) = \max \left( {\frac{{\gamma _{SIC}^{th}}}{{{a_c} - \gamma _{SIC}^{th}{a_t}}},\frac{z}{{{a_t}}}} \right)$ divides the expression into two items, such as $\gamma _t^{th} \in [0,{{a_t}{\Upsilon _1}}]$ and $\gamma _t^{th} \in [{{a_t}{\Upsilon _1}},\infty]$. Thus, substituting the Laplace transform expressions of the interference caused by other $RU$ links, this theorem is clarified via several mathematical manipulations.
\end{IEEEproof}
\end{theorem}

\begin{remark}
With the same analytical procedure of \textbf{Remark \ref{RT1}}, derivations in \textbf{Theorem \ref{ERRIS}} illustrate that the ergodic rate of typical user is proportional to the length of RISs $L$.
\end{remark}

\begin{corollary}\label{ERRISC1}
\emph{When fixing the path loss exponent of the typical user $\alpha_t = 2$, the closed-form ergodic rate expression of the typical user is derived as  }
\begin{align}
{{\mathbb{E}}}\left[ {R_t^{RIS}} \right] &= \frac{{\pi {\lambda _b}}}{{2\ln 2}}\sum\limits_{n = 1}^{{m_t}} {{{\left( { - 1} \right)}^{n + 1}} {
\binom{m_t}{n}
}} \notag\\
& \hspace*{0.3cm} \times \left( {\sum\limits_{j = 1}^J {{\omega _j}\sqrt {1 - \Xi  _j^2} } \frac{{{a_t}{\Upsilon _1}\left( {{\beta _{\rm{1}}}\left( {{\Upsilon _1}} \right)R_L^2{\rm{ + }}2{\beta _2}\left( {{\Upsilon _1}} \right)} \right)}}{{2\left( {1 + {\Xi _j}} \right)}}} \right.\notag\\
& \hspace*{0.3cm} \left. { + \sum\limits_{v = 1}^V {{\omega _v}\sqrt {1 - \Xi  _v^2} } \frac{{2{a_t}{\Upsilon _1}\left( {{\beta _{\rm{1}}}\left( {\frac{{{\Xi _v}}}{{{a_t}}}} \right)R_L^2{\rm{ + }}2{\beta _2}\left( {\frac{{{\Xi _v}}}{{{a_t}}}} \right)} \right)}}{{{{\left( {{\varpi _v} + 1} \right)}^2}\left( {1 + {\Xi _v}} \right)}}} \right),
\end{align}
\emph{where ${\Xi _j} = \frac{{{a_t}{\Upsilon _1}}}{2}\left( {{\varpi _j} + 1} \right)$, ${\varpi _j}{\rm{ = cos}}\left( {\frac{{2j - 1}}{{2J}}\pi } \right)$, ${\omega _j} = \pi /J$, ${\Xi _v} = \frac{{2{a_t}{\Upsilon _1}}}{{{\varpi _v} + 1}}$, ${\varpi _v}{\rm{ = cos}}\left( {\frac{{2j - 1}}{{2V}}\pi } \right)$ and ${\omega _v} = \pi/V $. }
\begin{IEEEproof}
Based on the closed-form expression of the coverage probability for the typical user in \textbf{Corollary \ref{CPRIS1}}, we are able to rewrite the ergodic rate expression as
\begin{align}
{\mathbb{E}}\left[ {R_t^{RIS}} \right] = \frac{{\pi {\lambda _b}}}{{2\ln 2}}\sum\limits_{n = 1}^{{m_t}} {{{\left( { - 1} \right)}^{n + 1}}{
\binom{m_c}{n}
} } \int_0^\infty  {\frac{{{\beta _{\rm{1}}}\left( {\Upsilon \left( z \right)} \right)R_L^2{\rm{ + }}2{\beta _2}\left( {\Upsilon \left( z \right)} \right)}}{{1 + z}}dz} .
\end{align}

Based on the derivations in \textbf{Theorem \ref{ERRIS}} and harnessing the Chebyshev-Gauss quadrature, we are able to derive the closed-form ergodic rate expression for the typical user.
\end{IEEEproof}
\end{corollary}

\begin{corollary}\label{ERRISC2}
\emph{When targeting the typical user's path loss exponent $\alpha_t = 4$, we derive the closed-form ergodic rate expression of the typical user as }
\begin{align}
{{\mathbb{E}}}\left[ {R_t^{RIS}} \right] &= \frac{1}{{\ln 2}}\sum\limits_{n = 1}^{{m_t}} {{{\left( { - 1} \right)}^{n + 1}}{
\binom{m_t}{n}
} } \sum\limits_{i = 1}^K {\sum\limits_{j = 1}^J {{\omega _i}{\omega _j}} \sqrt {1 - \Xi _i^2} \sqrt {1 - \Xi _j^2} } \notag\\
 &\hspace*{0.3cm}\times \frac{{{a_t}{\Upsilon _1}{\pi ^{\frac{3}{2}}}{\lambda _b}}}{{4R_L^{}\sqrt {{\beta _{\rm{1}}}\left( {{\Upsilon _1}} \right)} {\Xi _i}\left( {1 + z} \right)}}\exp \left( {\frac{{\beta _2^2\left( {{\Upsilon _1}} \right)}}{{4{\beta _{\rm{1}}}\left( {{\Upsilon _1}} \right)\Xi _i^4}}} \right)Erfc\left( {\frac{{\beta _2^{}\left( {{\Upsilon _1}} \right)}}{{2\sqrt {{\beta _{\rm{1}}}\left( {{\Upsilon _1}} \right)} \Xi _i^2}}} \right)\notag\\
 &\hspace*{0.3cm}+ \frac{1}{{\ln 2}}\sum\limits_{n = 1}^{{m_t}} {{{\left( { - 1} \right)}^{n + 1}}\binom{m_c}{n}} \sum\limits_{i = 1}^K {\sum\limits_{v = 1}^V {{\omega _i}{\omega _v}\sqrt {1 - \Xi _i^2} \sqrt {1 - \Xi _v^2} } }\notag \\
 &\hspace*{0.3cm}\times \frac{{{a_t}{\Upsilon _1}{\pi ^{\frac{3}{2}}}{\lambda _b}}}{{R_L^{}\sqrt {{\beta _{\rm{1}}}\left( {\frac{{{\Xi _v}}}{{{a_t}}}} \right)} {\Xi _i}{{\left( {1 + {\Xi _v}} \right)}^3}}}\exp \left( {\frac{{\beta _2^2\left( {\frac{{{\Xi _v}}}{{{a_t}}}} \right)}}{{4{\beta _{\rm{1}}}\left( {\frac{{{\Xi _v}}}{{{a_t}}}} \right)\Xi _i^4}}} \right)Erfc\left( {\frac{{\beta _2^{}\left( {\frac{{{\Xi _v}}}{{{a_t}}}} \right)}}{{2\sqrt {{\beta _{\rm{1}}}\left( {\frac{{{\Xi _v}}}{{{a_t}}}} \right)} \Xi _i^2}}} \right),
\end{align}
\emph{where the coefficients are the same as \textbf{Corollary \ref{ERRISC1}}.}
\begin{IEEEproof}
Substituting the closed-form coverage probability expression from \textbf{Corollary \ref{CPRIS2}} into \eqref{erdeft}, the ergodic rate expression is calculated as
\begin{align}
{{\mathbb{E}}}\left[ {R_t^{RIS}} \right] &=\frac{1}{{\ln 2}}\sum\limits_{n = 1}^{{m_t}} {{{\left( { - 1} \right)}^{n + 1}}
\binom{m_t}{n}
} \sum\limits_{i = 1}^K {\int_0^\infty  {\frac{{{\omega _i}{\pi ^{\frac{3}{2}}}{\lambda _b}\sqrt {1 - \Xi _i^2} }}{{2R_L^{}\sqrt {{\beta _{\rm{1}}}\left( {\Upsilon \left( z \right)} \right)} {\Xi _i}\left( {1 + z} \right)}}} }\notag \\
 &\hspace*{0.3cm}\times \exp \left( {\frac{{\beta _2^2\left( {\Upsilon \left( z \right)} \right)}}{{4{\beta _{\rm{1}}}\left( {\Upsilon \left( z \right)} \right)\Xi _i^4}}} \right)Erfc\left( {\frac{{\beta _2^{}\left( {\Upsilon \left( z \right)} \right)}}{{2\sqrt {{\beta _{\rm{1}}}\left( {\Upsilon \left( z \right)} \right)} \Xi _i^2}}} \right)dz,
\end{align}
and with the aid of Chebyshev-Gauss quadrature, the final expression is derived.
\end{IEEEproof}
\end{corollary}

\subsection{Ergodic Rate for the Connected User}
Recall that the distance from the nearest BS to the connected user is fixed. We also express the expression of ergodic rate via the coverage probability expression as
\begin{align}
{\mathbb{E}}\left[ {R_c^{RIS}} \right] = {\mathbb{E}}\left[ {{{\log }_2}\left( {1 + {\gamma _c}} \right)} \right] = \frac{1}{{\ln 2}}\int_0^\infty  {\frac{{{{\mathbb{P}}_c}\left( {\gamma _c^{th}} \right)}}{{1 + \gamma _c^{th}}}} d\gamma _c^{th},
\end{align}
and the approximated closed-form expression is derived via the following theorem.

\begin{theorem}\label{ERC}
\emph{Since we consider the RISs to enhance the channel environments of the typical user, there is a high probability that the channel conditions of the typical user are better than the connected users. Thus, the connected user is allocated at the first stage of SIC orders to escape from the SIC procedure. In this scenario, the ergodic rate of the connected user is derived as }
\begin{align}
{\mathbb{E}}\left[ {R_c^{RIS}} \right]  &= \frac{1}{{\ln 2}}\sum\limits_{n = 1}^{{m_c}} {{{\left( { - 1} \right)}^{n + 1}}{
\binom{m_c}{n}
} } \sum\limits_{w = 1}^W {{\omega _w}\sqrt {1 - \Xi _w^2} } \frac{{{\Upsilon _2}}}{{2\left( {1 + {\Xi _w}} \right)}}\notag\\
&\hspace*{0.3cm}\times{ \exp \left( { - {\mu _1}\left( {{\Xi _w}} \right)r_c^2 - {\mu _2}\left( {{\Xi _w}} \right){r_c}^{{\alpha _c}}} \right)} ,
\end{align}
\emph{where ${\Upsilon _2}={\frac{{{a_c}}}{{{a_t}}}}  $, ${\Xi _w} = \frac{{{\Upsilon _2}}}{2}\left( {{\varpi _w} + 1} \right)$, ${\varpi _w}{\rm{ = cos}}\left( {\frac{{2w - 1}}{{2W}}\pi } \right)$ and ${\omega _w} = \pi /W$.}
\begin{IEEEproof}
Based on the conditions when deriving the coverage probability, we obtain that $\gamma _c^{th}< \frac{{{a_c}}}{{{a_t}}}$. Therefore, the range of $\gamma _c^{th}$ is in $[0,{\Upsilon _2}]$. Based on the coverage probability expression of the connected user, we express the ergodic rate expression as
 \begin{align}
\mathbb{E}[R_c^{RIS}] &= \frac{1}{{\ln 2}}\sum\limits_{n = 1}^{{m_c}} {{{\left( { - 1} \right)}^{n + 1}}
\binom{m_c}{n}
} \int_0^{{\Upsilon _2}} {\frac{1}{{1 + z}}} {\exp \left( { - {\mu _1}\left( z \right)r_c^2 - {\mu _2}\left( z \right){r_c}^{{\alpha _c}}} \right)} dz,
 \end{align}
 and utilizing Chebyshev-Gauss quadrature, the proof is accomplished.
\end{IEEEproof}
\end{theorem}

\begin{remark}
According to the analytical procedure in \textbf{Remark \ref{RT2}}, we are able to conclude that the ergodic rate of the connected user is inversely proportional to the density of BSs $\lambda_b$.
\end{remark}

\subsection{Asymptotic Ergodic Rate for the Typical User}

In this subsection, we evaluate the coverage performance of the typical user when $L\to \infty$ holds. With the aid of the asymptotic expression of the exponential functions, we derive the asymptotic coverage probability and diversity gains.

\begin{corollary}\label{AERRIS}
\emph{We assume the half-length of the RIS to infinity, denoted as $L \to \infty$. With the aid of asymptotic expression, such as $\exp(-x)=1-x$ when $x \to \infty$, we derive the approximated expression of the ergodic rate for the GF user as }
\begin{align}
{\rm{E}}\left[ {R_t^{RIS}} \right]\left| {_{L \to \infty }} \right.& = \frac{{2\pi {\lambda _b}}}{{\ln 2R_L^2}}\int_0^\infty  {\int_0^{{R_L}} {\int_0^\infty  {\sum\limits_{n = 1}^{{m_t}} {{{\left( { - 1} \right)}^{n + 1}}
\binom{m_t}{n}} \frac{{2yx}}{{1 + z}}} } }\notag \\
 &\hspace*{0.3cm}\times \left( {1 - {\beta _{\rm{1}}}\left( {\Upsilon \left( z \right)} \right){{\left( {yx} \right)}^{{\alpha _t}}}} \right)\exp \left( { - {\beta _2}\left( {\Upsilon \left( z \right)} \right){x^2}} \right)dxdydz\notag\\
& = {C_1} - {C_2}f\left( {\frac{1}{{{L^2}}}} \right)
  \end{align}
  \emph{where $C_1$ and $C_2$ are constants and $f\left( {\frac{1}{{{L^2}}}} \right)$ represents a function negatively correlated to $L^2$.}
\begin{IEEEproof}
Substituting the asymptotic expression into the expression in \textbf{Theorem \ref{ERRIS}}, we achieve the asymptotic expression.
\end{IEEEproof}
\end{corollary}

\begin{remark}\label{Erslope}
\emph{When the condition $L \to \infty$ holds, which means we have a large RIS surface, we derive the slope to evaluate the performance as}
\begin{align}
S = \mathop {\lim }\limits_{L \to \infty } \frac{{{\rm{E}}\left[ {R_t^{RIS}} \right]\left| {_{L \to \infty }} \right.}}{{\log \left( L \right)}} = 0.
\end{align}
\end{remark}

\begin{remark}\label{Erslopeana}
Since the slope versus $L$ is zero, which represents that when we enhance the length of RISs, the performance would increase to an upper limit eventually. Although the channel conditions of the typical user would increase when we continuously enlarge the RISs, the interference from other BSs would also be enhanced, which leads to an upper limit.
\end{remark}

\section{Numerical Results}
In this section, numerical results are indicated to validate analytical and asymptotic expressions of coverage probability derived in the previous sections. We further accomplish the simulation results of the ergodic rate performance, including analytical and asymptotic expressions. Several comparisons are proposed to compare the performance under several cases, such as RIS-aided NOMA, RIS-aided OMA, and non-RIS-aided NOMA.

\subsection{Simulation Results on Coverage Probability}

In this subsection, numerical results validate analytical coverage probability for the typical user (\textbf{Theorem \ref{CPRIS}}) and the connected user (\textbf{Theorem \ref{CPC}}). Additionally, the asymptotic expressions via the length of RISs are validated to match the simulation results in high length region. Without otherwise specification, we define the numerical settings as: the half-length of RISs as $0.75$ m, the noise power as ${\sigma ^2} =  - 170 + 10\log \left( {f_c} \right) + {N_f}=-90 $ dB, where $f_c$ is the bandwidth as $10$ MHz and $N_f$ is noise figure as $10$ dB, transmit power of users as $\left[0,15\right]$ dBm, path loss exponents as $\alpha_c = 4$ and $\alpha_t = 2.4$, RIS ball radius $R_L = 25$ m, density of BSs as $\lambda_b = 1/(300^2\pi)$, thresholds $\gamma_{SIC}^{th}=\gamma_{t}^{th}=\gamma_c^{th}=10^{-2}$, Gamma distribution coefficient $m_c=m_t=4$ and power allocation coefficients $a_c = 0.6$ and $ a_t = 0.4$.

\subsubsection{Validation of Results on Coverage Probability}
\begin{figure}[!htb]
\centering
\includegraphics[width= 4in]{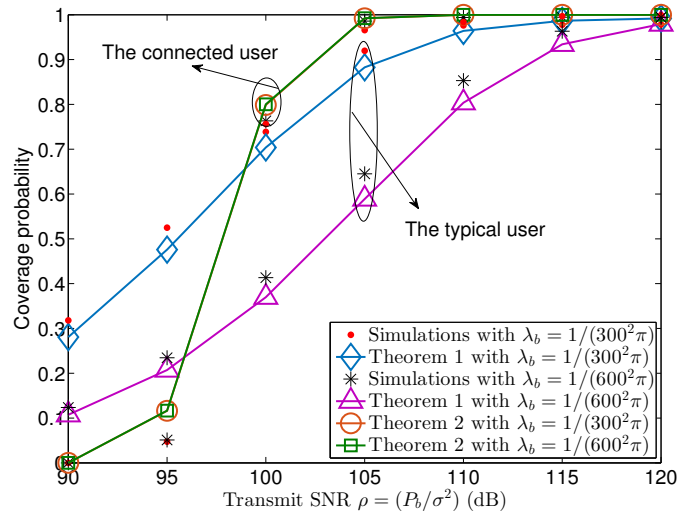}
\caption{Coverage probability versus transmit SNR with various density of BSs $\lambda_b=[1/(300^2\pi),1/(600^2\pi)]$: a verification of \textbf{Theorem \ref{CPRIS}} and \textbf{Theorem \ref{CPC}}. }
\label{figure1}
\end{figure}

The analytical coverage probability expressions of the typical user (\textbf{Theorem \ref{CPRIS}}) and the connected user (\textbf{Theorem \ref{CPC}}) are validated in Fig. \ref{figure1} . Additionally, we compare the performance of several scenarios with different density of BSs $\lambda_b$ in Fig. \ref{figure1}. The observation is that the density of BSs significantly affects the coverage performance of the typical user while it has a slight influence on connected users. This is because the total interference of the typical user is more sensitive than that of the connected users, thereby enhancing the density of BSs enlarges the interference of the typical user much more severely than the connected users.

\begin{figure}[!htb]
\centering
\includegraphics[width= 4in]{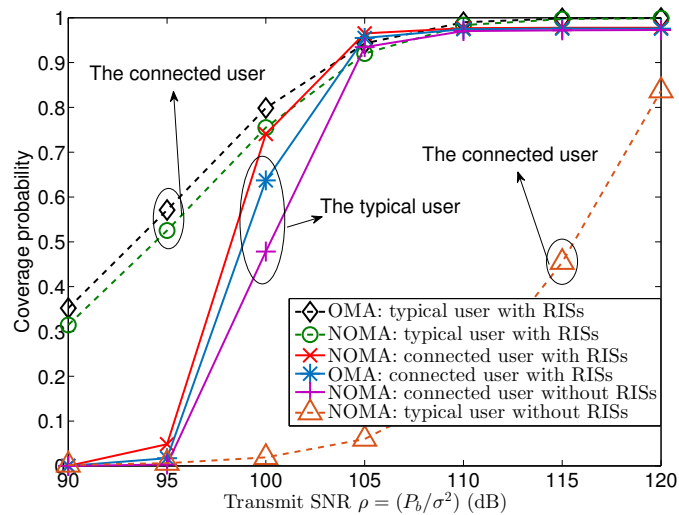}
\caption{Coverage probability versus transmit SNR: a comparison among conventional NOMA, RIS-aided OMA and RIS-aided NOMA scnearios.}
\label{figure2}
\end{figure}
\begin{figure}[!htb]
\centering
\includegraphics[width= 4in]{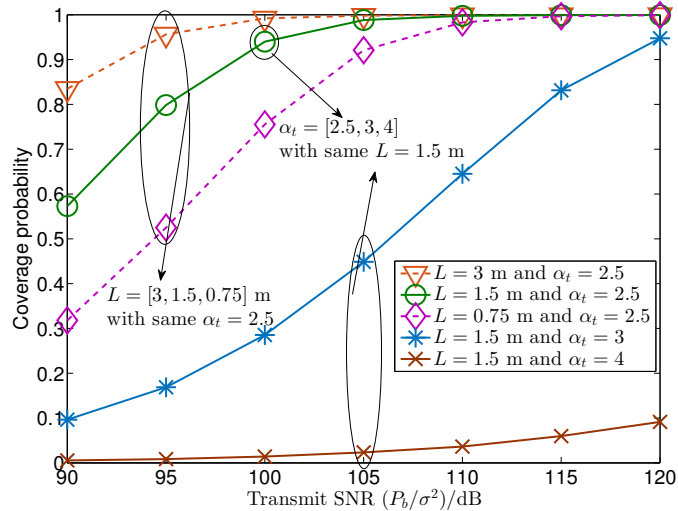}
\caption{Coverage probability versus transmit SNR with various half-length of RISs $L = [0.75, 1.5 ,3]$ m and path loss exponents $\alpha_t=[2.5,3,4]$.}
\label{figure3}
\end{figure}

\begin{figure}[!htb]
\centering
\includegraphics[width= 4in]{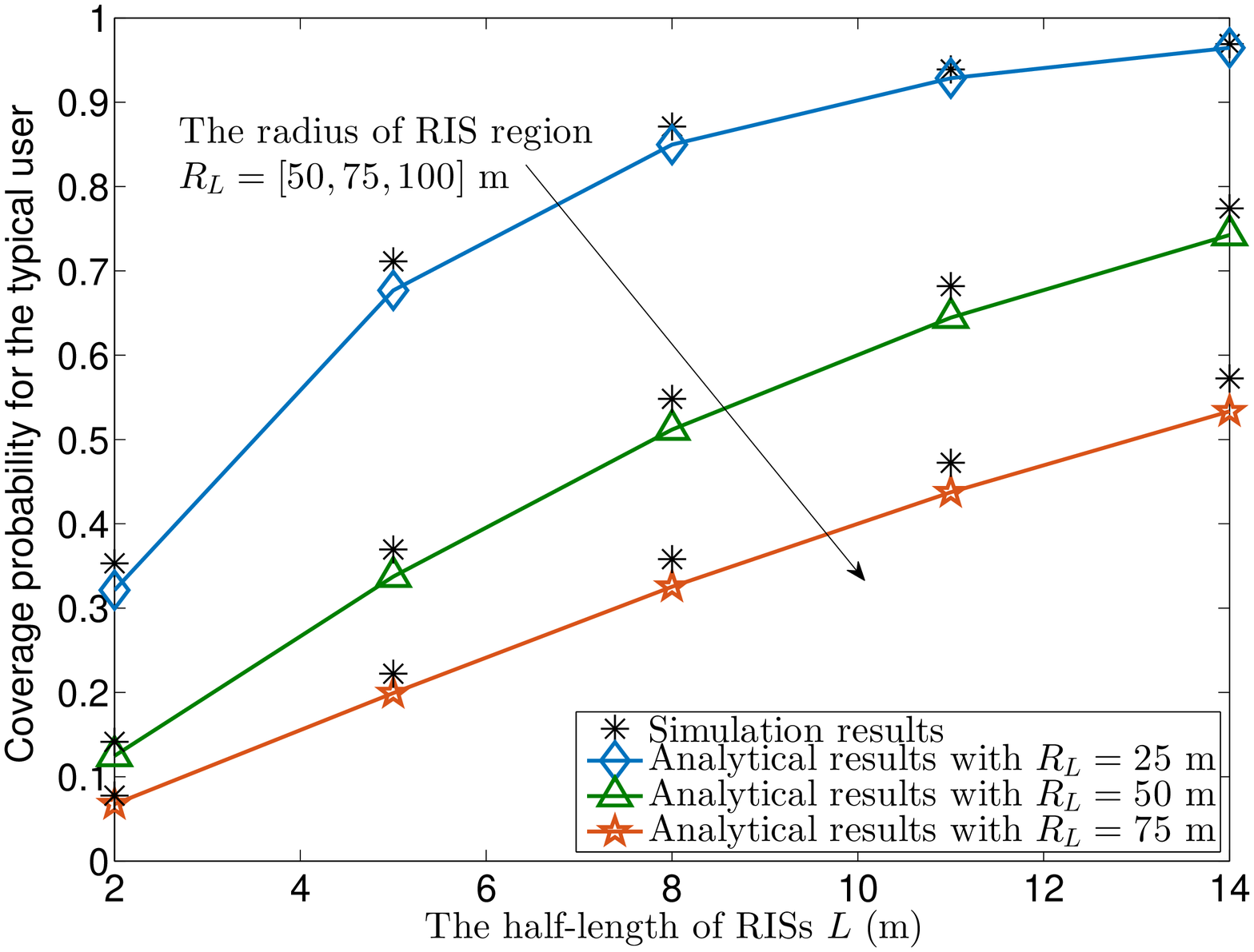}
\caption{Coverage probability versus the half-length of RISs $L$ m and path loss exponents $\alpha_t=[2.5,3,4]$.}
\label{figure4}
\end{figure}

\subsubsection{Performance Comparison}
The performance among conventional NOMA, RIS-aided NOMA, and RIS-aided OMA scenarios are compared in Fig. \ref{figure2}, which demonstrates that the performance of NOMA users boosts considerably with the aid of RISs, especially for the typical user. The enhancement of NOMA users are able to be explained that 1) when assisted with RISs, the connected users enable to avoid SIC procedures since the typical user with substantially improved channel gains takes over the SIC procedures, thereby the connected user would not experience outage scenarios caused by SIC failures; 2) with the aid of RISs, superior channel gains of the typical user increase coverage performance.

\begin{figure}[!htb]
\centering
\includegraphics[width= 4in]{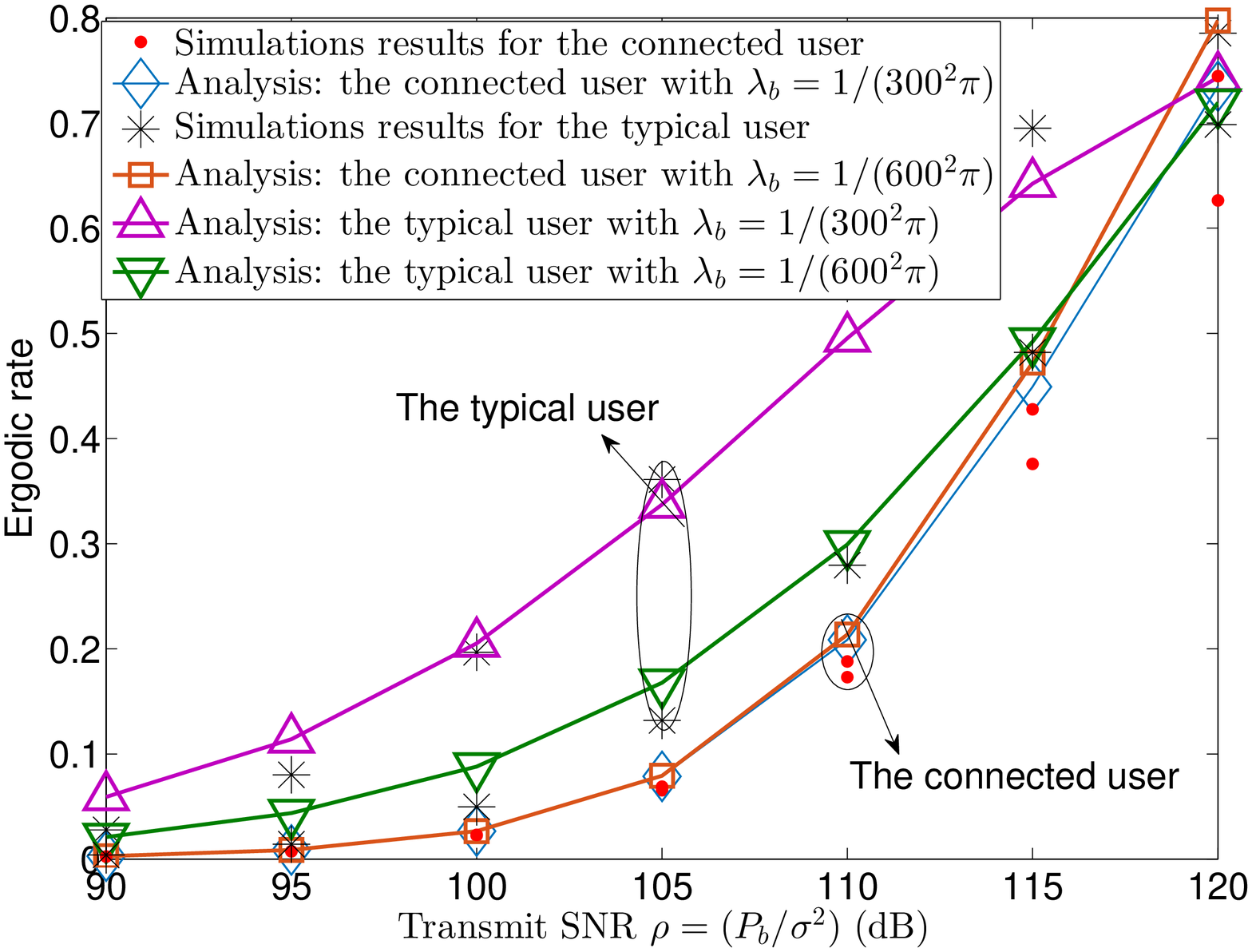}
\caption{Ergodic rates (BPCU) versus transmit SNR with various density values of BSs $\lambda_b = 1/(200^2\pi), 1/(400^2\pi), 1/(600^2\pi)$ for the typical user.}
\label{figure5}
\end{figure}

\begin{figure}[!htb]
\centering
\includegraphics[width= 4in]{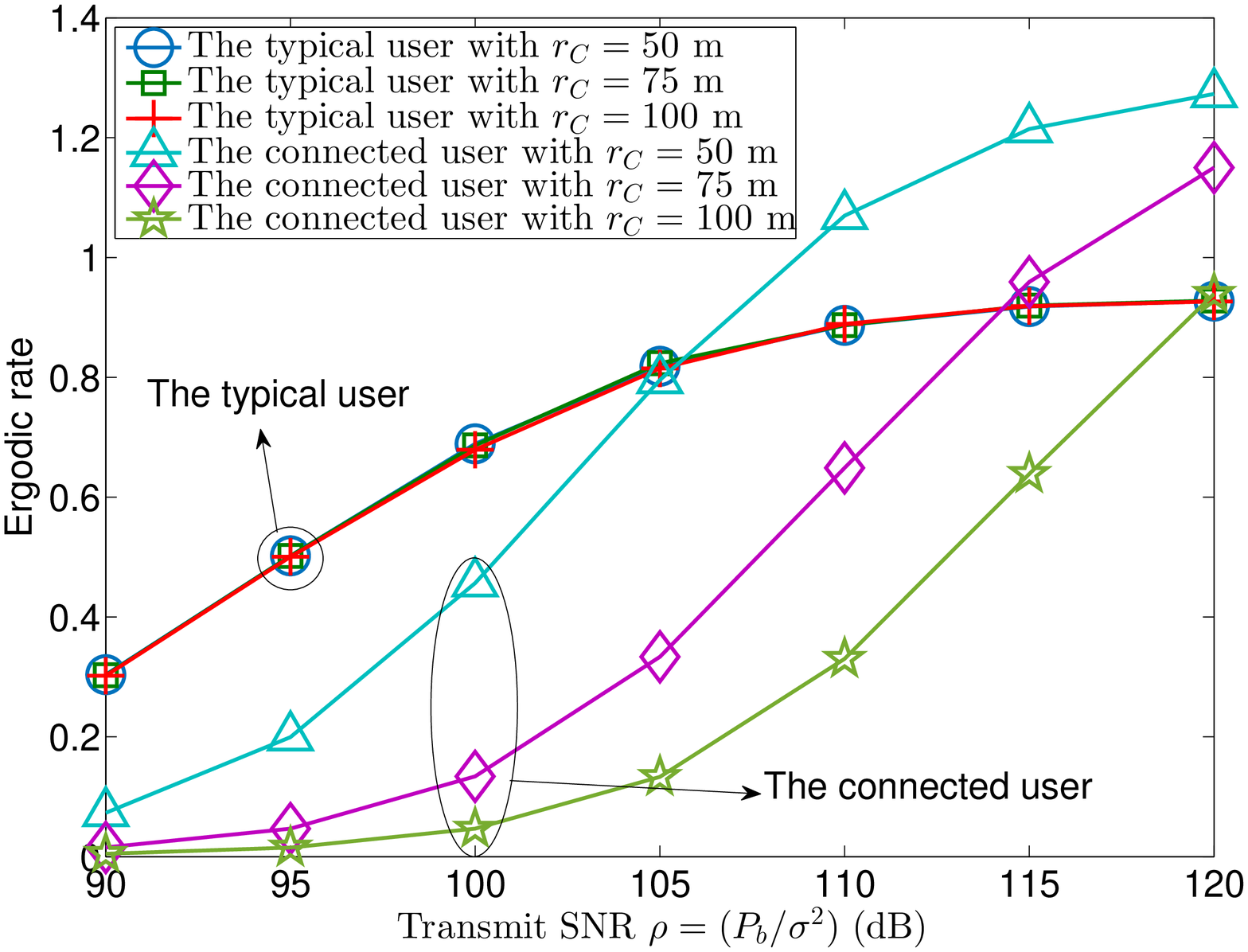}
\caption{Ergodic rates (BPCU) versus transmit SNR with various distances between the connected user and the BS as $r_C = 50,75,100$ m for the connected user.}
\label{figure6}
\end{figure}

\subsubsection{Outage Performance on Path Loss Exponents}
Considering the length of RISs, Fig. \ref{figure3} investigates the performance varied by $L$ and path loss exponents. Two observations are apparent to explain that: 1) long lengths of RISs cause high performance since more reconfigurable meta-material elements are involved and 2) enlarging path loss exponents results in reduced performance as the relationship between the path loss exponents and the coverage performance is a negative correlation.

\subsubsection{Outage Performance on the Length of RIS}
We investigate the outage performance versus the length of RIS via Fig. \ref{figure4}. It is observed that when we improve the length of RISs, the coverage probability is enlarged since the more signals and energy has been reflected via the bigger RISs than the small ones.

\subsection{Simulation Results on Ergodic Rage}
In this subsection, we validate analytical ergodic rate via numerical results for the typical user (\textbf{Theorem \ref{ERRIS}}) and connected users (\textbf{Theorem \ref{ERC}}). We note that the unit of ergodic rate is the bit per cell use (BPCU). We define the numerical settings as the same as the sections on coverage probability without otherwise specification.

\subsubsection{Validation of Ergodic Rate for the Typical User}
The ergodic rate performance with the unit BPCU versus the transmit SNR $P_b/\sigma^2$ is indicated in Fig. \ref{figure5}. We observe that when we have dense BSs, the performance of the typical user outperforms than users with small density. This is because a large density of BSs contributes to a short averaged distance between the nearest BS and the typical user, which leads to better performance than small density situations.

\subsubsection{Validation of Ergodic Rate for the connected User}
We validated the analytical results of the ergodic rate for the connected user via Fig. \ref{figure6}, which demonstrates the ergodic rate (BPCU) versus the transmit SNR. Observed by this figure, we conclude that when we enhance the distance between the connected user and the nearest BS, the performance goes worse as the long-distance leads to heavy path loss.

\section{Conclusion}

This paper has investigated the coverage probability and ergodic rate of RIS-assisted NOMA frameworks, where the PCP principle has been invoked to capture the spatial effects of NOMA users. The path loss models of RIS reflecting links has been derived, which is correlated with the ``product of distances" model to conform with long-distance regions. The angle distributions have been presented with a conclusion that the BS-RIS-User angles obey uniform distributions in $[0.\pi]$. With the aid of the derived closed-form expressions of coverage probabilities and numerical results, the performance of conventional NOMA, RIS-aided NOMA, and RIS-aided OMA scenarios have been compared, which has shown that RISs enhance the performance of users significantly. The asymptotic expressions of ergodic rates for the typical user have illustrated that the performance has upper limits when enhancing the length of the RIS. The analysis of this paper has verified that two applications of RISs in multi-cell NOMA networks are feasible, such that: 1) RISs are able to improve the channel conditions of blocked or far users; 2) RISs enable to alter the SIC order to maintain primary users avoiding SIC procedures.

\section*{Appendix~A: Proof of Lemma~\ref{I_c}} \label{Appendix:A}
\renewcommand{\theequation}{A.\arabic{equation}}
\setcounter{equation}{0}
With the aid of the expansion of the exponential function and based on \eqref{I_C_def}, we are able to express the Laplace transform of the interference for the connected user as
\begin{align}\label{A1}
{\cal L}_{(s)}^c& =\mathbb{E} \left[ {\exp \left( { - \sum\limits_{{{\bf{x}}_I} \in {\Phi _r} \setminus {{\bf{x}}_B}} {s{P_b}C{{\left| {{h_{t,{{\bf{x}}_I}}}} \right|}^2}{r_{c,I}}^{ - {\alpha _c}}} } \right)} \right]\notag\\
&= \mathbb{E}\left[ {\prod\limits_{{{\bf{x}}_I} \in {\Phi _r} \setminus {{\bf{x}}_B}} {\exp } \left( { - s{P_b}C{{\left| {{h_{t,{{\bf{x}}_I}}}} \right|}^2}{r_{c,I}}^{ - {\alpha _c}}} \right)} \right].
\end{align}

Based on the binomial expansion, we obtain a tight upper bound for the normalized Gamma variables $\left|h\right|^2$, denoted as $\mathbb{P}\left\{\left|h\right|^2<x\right\}= \left(1-\exp(-x\eta )\right)^m$, where $m$ is the coefficient of Gamma distribution and $\eta = m(m!)^{-\frac{1}{m}}$. Thus, the equation \eqref{A1} is approximated expressed as
\begin{align}
{\cal L}_{(s)}^c = \mathbb{E}\left[ {\prod\limits_{{{\bf{x}}_I} \in {\Phi _r} \setminus {{\bf{x}}_B}} {{{\left( {1 + \frac{{s{P_b}C{r_{c,I}}^{ - {\alpha _c}}}}{m}} \right)}^{ - m}}} } \right].
\end{align}

For some stochastic processes, the probability generating functional (PGFL) are useful tools for dealing with sums and limits of
random variables, which is expressed as
\begin{align}
\mathbb{E}\left[ {\prod\limits_{x \in \Phi } {f\left( x \right)} } \right] = \exp \left( { - \lambda \int\limits_{{\mathbb{R}^2}} {\left( {1 - f\left( x \right)} \right)dx} } \right),
\end{align}
where $\Phi$ is a PPP with density $\lambda$ and $f\left( x \right):{R^2} \to \left[ {0,1} \right]$ is a real value function. Then, we utilize PGFL to derive the Laplace transform as
\begin{align}
{\cal L}_{(s)}^c = \exp \left( { - 2\pi {\lambda _b}\int_{{r_c}}^\infty  {\left( {1 - \left( {{{\left( {1 + \frac{{s{P_b}C{r^{ - {\alpha _c}}}}}{m}} \right)}^{ - m}}} \right)} \right)rdr} } \right),
\end{align}
and via the integration $\int_A^\infty  {\left( {1 - {{\left( {1 + s{y^{ - \alpha }}} \right)}^{ - N}}} \right)ydy = \frac{{{A^2}}}{2}\left( {{}_2{F_1}\left( { - \frac{2}{\alpha },N;1 - \frac{2}{\alpha }; - \frac{s}{{{A^\alpha }}}} \right) - 1} \right)} $, this lemma is proved.

\section*{Appendix~B: Proof of Lemma~\ref{I_t}} \label{Appendix:B}
\renewcommand{\theequation}{B.\arabic{equation}}
\setcounter{equation}{0}
With the aid of the properties of exponential functions such as $e^{ab} = e^a\cdot e^b$, the Laplace transform expressions of the interference from other cells for the typical user is expressed as
\begin{align}\label{B1}
{\cal L}_{(s)}^{t,RIS}& = \mathbb{E}\left[ {\exp \left( { - {\rho _t}\sum\limits_{{{\bf{x}}_I} \in {\Phi _r} \setminus {{\bf{x}}_B}} {\frac{{s{P_b}C_{RIS,I}^2{{\left| {{h_{t,{{\bf{x}}_I}}}} \right|}^2}}}{{{{\left( {{r_{BR\left( I \right)}}{r_{RU\left( I \right)}}} \right)}^{{\alpha _t}}}}}} } \right)} \right]\notag\\
 &= \mathbb{E}\left[ {\prod\limits_{{{\bf{x}}_I} \in {\Phi _r} \setminus {{\bf{x}}_B}} {{{\left( {1 + \frac{{s{P_b}C_{RIS,I}^2}}{{2m{{\left( {{r_{BR\left( I \right)}}{r_{RU\left( I \right)}}} \right)}^{{\alpha _t}}}}}} \right)}^{ - m}}} } \right].
\end{align}

Based on the PGFL theorem, the Laplace transform \eqref{B1} is derived as
\begin{align}
{\cal L}_{(s)}^{t,RIS}& =  \exp \left( { - 2\pi {\lambda _b}\int_{{r_{BR\left( 0 \right)}}}^\infty  {\left( {1 - {{\left( {1 + \frac{{s{P_b}C_{RIS,E}^2}}{{m{{\left( {x{r_{RU\left( 0 \right)}}} \right)}^{{\alpha _t}}}}}} \right)}^{ - m}}} \right)xdx} } \right).
\end{align}

Exploiting the integration expressions in Appendix~A, the lemma is proved.

\section*{Appendix~C: Proof of Theorem~\ref{CPRIS}} \label{Appendix:C}
\renewcommand{\theequation}{C.\arabic{equation}}
\setcounter{equation}{0}
The Campbell's theorem is expressed as
$\mathbb{E}\left[ {\sum\limits_{x \in \Phi }^{} {f\left( x \right)} } \right] = \lambda \int\limits_{{\mathbb{R}^2}} {f\left( x \right)dx}$,
where $\Phi$ is a PPP with density $\lambda$. Based on the Campbell's theorem, the expressions of the average interference for the connected user is derived as
\begin{align}
\mathbb{E}\left[ {{I_c}} \right] = 2\pi {\lambda _b}\int_{{r_c}}^\infty  {{P_b}C{r^{1 - {\alpha _c}}}} dr{\rm{ = }}\frac{{2\pi {\lambda _b}{P_b}Cr_c^{2 - {\alpha _c}}}}{{{\alpha _c} - 2}}.
\end{align}

Note that the normalized Gamma variables have a tight upper bound, denoted as $\mathbb{P}\left[\left|h^2\right|<x\right] <\left(1-e^{-x{\eta _t}}\right)^{m_t}$. Utilizing binomial expansions, the expressions of coverage probability for the typical user is expressed as
\begin{align}\label{C3}
{\mathbb{P}}_{t} \approx  \sum\limits_{n = 1}^{{m_t}} {{{\left( { - 1} \right)}^{n + 1}}{\binom{m_t}{n}}} {\cal L}\left[ {\frac{{n{\eta _t}\Upsilon {I_{t,{\rho _t}}}}}{{{P_b}P_t^{RIS}}}} \right]\mathbb{E}\left[ {{e^{ - \frac{{n{\eta _t}\Upsilon {\sigma ^2}}}{{{P_b}P_t^{RIS}}}}}} \right],
\end{align}
and via substituting \textbf{Lemma \ref{EverageAngle}} and \textbf{Lemma \ref{I_t}} into the equation above, the theorem is verified.


\bibliographystyle{IEEEtran}
\bibliography{mybib}

\begin{thebibliography}{10}
\providecommand{\url}[1]{#1}
\csname url@samestyle\endcsname
\providecommand{\newblock}{\relax}
\providecommand{\bibinfo}[2]{#2}
\providecommand{\BIBentrySTDinterwordspacing}{\spaceskip=0pt\relax}
\providecommand{\BIBentryALTinterwordstretchfactor}{4}
\providecommand{\BIBentryALTinterwordspacing}{\spaceskip=\fontdimen2\font plus
\BIBentryALTinterwordstretchfactor\fontdimen3\font minus
  \fontdimen4\font\relax}
\providecommand{\BIBforeignlanguage}[2]{{%
\expandafter\ifx\csname l@#1\endcsname\relax
\typeout{** WARNING: IEEEtran.bst: No hyphenation pattern has been}%
\typeout{** loaded for the language `#1'. Using the pattern for}%
\typeout{** the default language instead.}%
\else
\language=\csname l@#1\endcsname
\fi
#2}}
\providecommand{\BIBdecl}{\relax}
\BIBdecl

\bibitem{YuanweiNOMA}
Y.~{Liu}, Z.~{Qin}, M.~{Elkashlan}, Z.~{Ding}, A.~{Nallanathan}, and
  L.~{Hanzo}, ``Nonorthogonal multiple access for 5{G} and beyond,''
  \emph{Proc. IEEE}, vol. 105, no.~12, pp. 2347--2381, 2017.

\bibitem{9086766}
M.~A. {El Mossallamy}, H.~{Zhang}, L.~{Song}, K.~G. {Seddik}, Z.~{Han}, and
  G.~Y. {Li}, ``Reconfigurable intelligent surfaces for wireless
  communications: Principles, challenges, and opportunities,'' \emph{IEEE
  Trans. Cogn. Commun. Netw.}, doi: 10.1109/TCCN.2020.2992604, 2020.

\bibitem{8910627}
Q.~{Wu} and R.~{Zhang}, ``Towards smart and reconfigurable environment:
  Intelligent reflecting surface aided wireless network,'' \emph{{IEEE} Commun.
  Mag.}, vol.~58, no.~1, pp. 106--112, Jan. 2020.

\bibitem{RISURSorRIS}
M.~{Di Renzo}, F.~{Habibi Danufane}, X.~{Xi}, J.~{de Rosny}, and
  S.~{Tretyakov}, ``Analytical modeling of the path-loss for reconfigurable
  intelligent surfaces$-$anomalous mirror or scatterer?'' \emph{Proc. IEEE 21th
  Int. Workshop Signal Process. Adv. Wireless Commun. (SPAWC)}, pp. 1--5, 2020.

\bibitem{Passlossmodel}
W.~{Tang}, M.~Z. {Chen}, X.~{Chen}, J.~Y. {Dai}, Y.~{Han}, M.~D. {Renzo},
  Y.~{Zeng}, S.~{Jin}, Q.~{Cheng}, and T.~J. {Cui}, ``Wireless communications
  with reconfigurable intelligent surface: Path loss modeling and experimental
  measurement,'' \emph{arXiv preprint arXiv:1911.05326}, 2019.

\bibitem{wirelesscom}
E.~{Basar}, M.~{Di Renzo}, J.~{De Rosny}, M.~{Debbah}, M.~{Alouini}, and
  R.~{Zhang}, ``Wireless communications through reconfigurable intelligent
  surfaces,'' \emph{IEEE Access}, vol.~7, pp. 116\,753--116\,773, 2019.

\bibitem{pan}
G.~{Pan}, J.~{Ye}, J.~{An}, and M.~S. {Alouini}, ``When full-duplex
  transmission meets intelligent reflecting surface: Opportunities and
  challenges,'' \emph{arXiv preprint arXiv:2005.12561}, 2019.

\bibitem{7982794}
Y.~{Liu}, Z.~{Qin}, M.~{Elkashlan}, A.~{Nallanathan}, and J.~A. {McCann},
  ``Non-orthogonal multiple access in large-scale heterogeneous networks,''
  \emph{{IEEE} J. Sel. Areas Commun.}, vol.~35, no.~12, pp. 2667--2680, Dec.
  2017.

\bibitem{8856258}
W.~{Yi}, Y.~{Liu}, E.~{Bodanese}, A.~{Nallanathan}, and G.~K. {Karagiannidis},
  ``A unified spatial framework for {UAV}-aided mm{W}ave networks,''
  \emph{{IEEE} Trans. Commun.}, vol.~67, no.~12, pp. 8801--8817, 2019.

\bibitem{GammaDistributionTransfer}
T.~{Bai} and R.~W. {Heath}, ``Coverage and rate analysis for millimeter-wave
  cellular networks,'' \emph{{IEEE} Trans. Wireless Commun.}, vol.~14, no.~2,
  pp. 1100--1114, 2015.

\bibitem{haenggi}
M.~{Haenggi}, ``On distances in uniformly random networks,'' \emph{IEEE Trans.
  Inf. Theory}, vol.~51, no.~10, pp. 3584--3586, 2005.

\bibitem{Beamforming1}
W.~{Yan}, X.~{Yuan}, Z.~{He}, and X.~{Kuai}, ``Passive beamforming and
  information transfer design for reconfigurable intelligent surfaces aided
  multiuser {MIMO} systems,'' \emph{{IEEE} J. Sel. Areas Commun.}, doi:
  10.1109/JSAC.2020.3000811, 2020.

\bibitem{Beamforming2}
M.~{Jung}, W.~{Saad}, M.~{Debbah}, and C.~S. {Hong}, ``On the optimality of
  reconfigurable intelligent surfaces ({RIS}s): Passive beamforming,
  modulation, and resource allocation,'' \emph{arXiv preprint
  arXiv:1910.00968}, 2019.

\bibitem{Beamforming3}
B.~{Di}, H.~{Zhang}, L.~{Song}, Y.~{Li}, Z.~{Han}, and H.~V. {Poor}, ``Hybrid
  beamforming for reconfigurable intelligent surface based multi-user
  communications: Achievable rates with limited discrete phase shifts,''
  \emph{{IEEE} J. Sel. Areas Commun.}, doi: 10.1109/JSAC.2020.3000813, 2020.

\bibitem{OP1}
J.~{Ye}, S.~{Guo}, and M.~{Alouini}, ``Joint reflecting and precoding designs
  for ser minimization in reconfigurable intelligent surfaces assisted {MIMO}
  systems,'' \emph{{IEEE} Trans. Wireless Commun.}, vol.~19, no.~8, pp.
  5561--5574, 2020.

\bibitem{OP2}
H.~{Guo}, Y.~{Liang}, J.~{Chen}, and E.~G. {Larsson}, ``Weighted sum-rate
  maximization for reconfigurable intelligent surface aided wireless
  networks,'' \emph{{IEEE} Trans. Wireless Commun.}, vol.~19, no.~5, pp.
  3064--3076, 2020.

\bibitem{OP3}
M.~{Jung}, W.~{Saad}, M.~{Debbah}, and C.~S. {Hong}, ``{MIMO-NOMA} networks
  relying on reconfigurable intelligent surface: A signal cancellation based
  design,'' \emph{arXiv preprint arXiv:1910.00968}, 2019.

\bibitem{Deeplearning1}
C.~{Huang}, R.~{Mo}, and C.~{Yuen}, ``Reconfigurable intelligent surface
  assisted multiuser miso systems exploiting deep reinforcement learning,''
  \emph{{IEEE} J. Sel. Areas Commun.}, doi: 10.1109/JSAC.2020.3000835, 2020.

\bibitem{Deeplearning2}
S.~{Khan} and S.~Y. {Shin}, ``Deep-learning-aided detection for reconfigurable
  intelligent surfaces,'' \emph{arXiv preprint arXiv:1910.09136}, 2019.

\bibitem{scenario1}
X.~{Yang}, C.~{Wen}, and S.~{Jin}, ``{MIMO} detection for reconfigurable
  intelligent surface-assisted millimeter wave systems,'' \emph{{IEEE} J. Sel.
  Areas Commun.}, doi: 10.1109/JSAC.2020.3000822, 2020.

\bibitem{scenario2}
N.~S. {Perovi\'{c}}, M.~D. {Renzo}, and M.~F. {Flanagan}, ``Channel capacity
  optimization using reconfigurable intelligent surfaces in indoor mm{W}ave
  environments,'' \emph{Proc. IEEE Int. Conf. Commun. ({ICC})}, pp. 1--7, 2020.

\bibitem{scenario3}
A.~U. {Makarfi}, K.~M. {Rabie}, O.~{Kaiwartya}, O.~S. {Badarneh}, X.~{Li}, and
  R.~{Kharel}, ``Reconfigurable intelligent surface enabled {IoT} networks in
  generalized fading channels,'' \emph{Proc. IEEE Int. Conf. Commun. ({ICC})},
  pp. 1--6, 2020.

\bibitem{scenario4}
A.~{Khaleel} and E.~{Basar}, ``Reconfigurable intelligent surface-empowered
  {MIMO} systems,'' \emph{IEEE Syst. J.}, pp. 1--9, 2020.

\bibitem{sumrate}
M.~{Zeng}, X.~{Li}, G.~{Li}, W.~{Hao}, and O.~A. {Dobre}, ``Sum rate
  maximization for {IRS}-assisted uplink {NOMA},'' \emph{arXiv preprint
  arXiv:2004.10791}, 2020.

\bibitem{BER}
V.~C. {Thirumavalavan} and T.~S. {Jayaraman}, ``{BER} analysis of
  reconfigurable intelligent surface assisted downlink power domain {NOMA}
  system,'' \emph{Proc. Int. Conf. Commun. Syst. Netw. (COMSNETS)}, pp.
  519--522, 2020.

\bibitem{power2}
J.~{Zhu}, Y.~{Huang}, J.~{Wang}, K.~{Navaie}, and Z.~{Ding}, ``Power efficient
  {IRS}-assisted {NOMA},'' \emph{arXiv preprint arXiv:1912.11768}, 2019.

\bibitem{power1}
B.~{Zheng}, Q.~{Wu}, and R.~{Zhang}, ``Intelligent reflecting surface-assisted
  multiple access with user pairing: {NOMA} or {OMA}?'' \emph{{IEEE} Commun.
  Lett.}, vol.~24, no.~4, pp. 753--757, 2020.

\bibitem{power3}
M.~{Fu}, Y.~{Zhou}, Y.~{Shi}, and K.~B. {Letaief}, ``Reconfigurable intelligent
  surface empowered downlink non-orthogonal multiple access,'' \emph{arXiv
  preprint arXiv:1910.07361}, 2019.

\bibitem{zhiguo2}
Z.~{Ding} and H.~{Vincent Poor}, ``A simple design of {IRS}-{NOMA}
  transmission,'' \emph{IEEE Commun. Lett.}, vol.~24, no.~5, pp. 1119--1123,
  2020.

\bibitem{zhiguo1}
Z.~{Ding}, R.~{Schober}, and H.~V. {Poor}, ``On the impact of phase shifting
  designs on {IRS}-{NOMA},'' \emph{IEEE Wireless Commun. Lett.}, doi:
  10.1109/LWC.2020.2991116, 2020.

\bibitem{tianwei1}
T.~{Hou}, Y.~{Liu}, Z.~{Song}, X.~{Sun}, Y.~{Chen}, and L.~{Hanzo},
  ``Reconfigurable intelligent surface aided {NOMA} networks,'' \emph{{IEEE} J.
  Sel. Areas Commun.}, doi: 10.1109/JSAC.2020.3007039, 2020.

\bibitem{yanyu}
Y.~{Cheng}, K.~H. {Li}, Y.~{Liu}, K.~C. {Teh}, and H.~V. {Poor}, ``Downlink and
  uplink intelligent reflecting surface aided networks: {NOMA} and {OMA},''
  \emph{arXiv preprint arXiv:2005.00996}, 2019.

\bibitem{tianwei2}
T.~{Hou}, Y.~{Liu}, Z.~{Song}, X.~{Sun}, Y.~{Chen}, and L.~{Hanzo}, ``{MIMO}
  assisted networks relying on large intelligent surfaces: A stochastic
  geometry model,'' \emph{arXiv preprint arXiv:1910.00959}, 2019.

\bibitem{8401954}
W.~{Yi}, Y.~{Liu}, and A.~{Nallanathan}, ``Cache-enabled {HetNets} with
  millimeter wave small cells,'' \emph{{IEEE} Trans. Commun.}, vol.~66, no.~11,
  pp. 5497--5511, Nov. 2018.

\bibitem{di2019reflection}
M.~D.~Renzo and J.~Song, ``Reflection probability in wireless networks with
  metasurface-coated environmental objects: an approach based on random spatial
  processes,'' \emph{EURASIP J. on Wireless Commun. and Netw.}, vol. 2019,
  no.~1, p.~99, 2019.

\bibitem{ntontin2019reconfigurable}
M.~{Di Renzo}, K.~{Ntontin}, J.~{Song}, F.~H. {Danufane}, X.~{Qian},
  F.~{Lazarakis}, J.~{De Rosny}, D.~{Phan-Huy}, O.~{Simeone}, R.~{Zhang},
  M.~{Debbah}, G.~{Lerosey}, M.~{Fink}, S.~{Tretyakov}, and S.~{Shamai},
  ``Reconfigurable intelligent surfaces vs. relaying: Differences,
  similarities, and performance comparison,'' \emph{IEEE Open J. Commun. Soc.},
  vol.~1, pp. 798--807, 2020.

\bibitem{8876629}
W.~{Yi}, Y.~{Liu}, Y.~{Deng}, A.~{Nallanathan}, and R.~W. {Heath}, ``Modeling
  and analysis of mm{W}ave {V2X} networks with vehicular platoon systems,''
  \emph{{IEEE} J. Sel. Areas Commun.}, vol.~37, no.~12, pp. 2851--2866, 2019.

\bibitem{yuanwei1}
Y.~{Liu}, Z.~{Qin}, M.~{Elkashlan}, Y.~{Gao}, and L.~{Hanzo}, ``Enhancing the
  physical layer security of non-orthogonal multiple access in large-scale
  networks,'' \emph{{IEEE} Trans. Wireless Commun.}, vol.~16, no.~3, pp.
  1656--1672, 2017.

\bibitem{nNearestUser}
{D. {Moltchanov}}, ``Distance distributions in random networks,'' \emph{Ad Hoc
  Networks}, vol.~10, no.~6, pp. 1146--1166, 2012.

\bibitem{nNearestUserJ}
J.~G. {Andrews}, F.~{Baccelli}, and R.~K. {Ganti}, ``A tractable approach to
  coverage and rate in cellular networks,'' \emph{{IEEE} Trans. Commun.},
  vol.~59, no.~11, pp. 3122--3134, 2011.

\bibitem{table}
A.~P. {Prudnikov}, Y.~A. {Brychkov}, and O.~I. {Marichev}, ``\emph{Integrals
  and series, vol. 1, special functions},'' 1986.

\end{thebibliography}
\end{document}